\documentclass[usenatbib]{mn2e} 
\usepackage{graphicx,fixltx2e,amsmath,amssymb,amsfonts,latexsym,wasysym}

\def\chisqr{\hbox{$\chi^2_{\rm r}$}}
\def\msun{\hbox{${\rm M}_{\odot}$}}

\def\mspy{\hbox{${\rm M}_{\odot}$\,yr$^{-1}$}}
\def\rsun{\hbox{${\rm R}_{\odot}$}}
\def\lsun{\hbox{${\rm L}_{\odot}$}}
\def\rcor{\hbox{$r_{\rm cor}$}}
\def\rmag{\hbox{$r_{\rm mag}$}}

\def\rstar{\hbox{$R_{\star}$}}

\def\teff{\hbox{$T_{\rm eff}$}}
\def\logg{\hbox{$\log g$}}
\def\sn{\hbox{S/N}}

\def\kms{\hbox{km\,s$^{-1}$}}
\def\vsini{\hbox{$v \sin i$}}

\def\ptt{\hbox{$10^{-4} I_{\rm c}$}}
\def\mV{\hbox{$m_{\rm V}$}}

\def\degr{\hbox{$^\circ$}}

\newcommand{\caii}{\hbox{Ca$\;${\sc ii}}}
\newcommand{\fei}{\hbox{Fe$\;${\sc i}}}

\newcommand{\hei}{\hbox{He$\;${\sc i}}}
\newcommand{\hal}{\hbox{H${\alpha}$}}
\newcommand{\hbe}{\hbox{H${\beta}$}}

\begin{document}

\title[Magnetism \& accretion of TW~Hya]{The large-scale magnetic field and poleward mass accretion of the 
classical T~Tauri star TW~Hya}  
\makeatletter

\def\newauthor{%
  \end{author@tabular}\par
  \begin{author@tabular}[t]{@{}l@{}}}
\makeatother
 
\author[J.-F.~Donati et al.]
{\vspace{1.7mm}
J.-F.~Donati$^1$\thanks{E-mail: 
donati@ast.obs-mip.fr }, 
S.G.~Gregory$^2$, S.H.P.~Alencar$^3$, J.~Bouvier$^4$, G.~Hussain$^5$, \\  
\vspace{1.7mm}
{\hspace{-1.5mm}\LARGE\rm 
M.~Skelly$^1$, C.~Dougados$^4$, M.M.~Jardine$^6$, F.~M\'enard$^4$, M.M.~Romanova$^7$ }\\ 
\vspace{1.7mm}
{\hspace{-1.5mm}\LARGE\rm
Y.C.~Unruh$^8$ \& the MaPP collaboration} \\
$^1$ IRAP--UMR 5277, CNRS \& Univ.\ de Toulouse, 14 Av.\ E.~Belin, F--31400 Toulouse, France \\
$^2$ California Institute of Technology, MC 249-17, Pasadena, CA 91125, USA \\ 
$^3$ Departamento de F\`{\i}sica -- ICEx -- UFMG, Av. Ant\^onio Carlos, 6627, 30270-901 Belo Horizonte, MG, Brazil \\ 
$^4$ IPAG--UMR 5274, CNRS \& Univ.\ J.~Fourier, 414 rue de la Piscine, F--38041 Grenoble, France \\ 
$^5$ ESO, Karl-Schwarzschild-Str.\ 2, D-85748 Garching, Germany \\ 
$^6$ School of Physics and Astronomy, Univ.\ of St~Andrews, St~Andrews, Scotland KY16 9SS, UK \\ 
$^7$ Department of Astronomy, Cornell University, Ithaca, NY 14853-6801, USA \\ 
$^8$ Astrophysics Group, Blackett Laboratory, Imperial College London, SW7 2AZ, UK
}

\date{2011 June, MNRAS in press}
\maketitle
 
\begin{abstract}  

We report here results of spectropolarimetric observations of the $\simeq$8~Myr classical 
T~Tauri star (cTTS) TW~Hya carried out with ESPaDOnS at the Canada-France-Hawaii Telescope (CFHT) 
in the framework of the `Magnetic Protostars and Planets' (MaPP) programme, and obtained at 2 
different epochs (2008 March and 2010 March).  Obvious Zeeman signatures are detected at all 
times, both in photospheric lines and in accretion-powered emission lines.  Significant intrinsic 
variability and moderate rotational modulation is observed in both photospheric and accretion 
proxies. 

Using tomographic imaging, we reconstruct maps of the large-scale field, of the photospheric 
brightness and of the accretion-powered emission at the surface of TW~Hya at both epochs.  
We find that the magnetic topology is mostly poloidal and axisymmetric with respect to the rotation 
axis of the star, and that the octupolar component of the large-scale field (2.5--2.8~kG 
at the pole) largely dominates the dipolar component.  This large-scale 
field topology is characteristic of partly-convective stars, supporting the conclusion (from 
evolutionary models) that TW~Hya already hosts a radiative core.  We also show that TW~Hya features 
a high-latitude photospheric cool spot overlapping with the main magnetic pole (and producing 
the observed radial velocity fluctuations);  this is also where accretion concentrates most of 
the time, although accretion at lower latitudes is found to occur episodically.  

We propose that the relatively rapid rotation of TW~Hya (with respect to AA~Tau-like cTTSs) 
directly reflects the weakness of the large-scale dipole, no longer capable of magnetically 
disrupting the accretion disc up to the corotation radius (at which the Keplerian period 
equals the stellar rotation period).  We therefore conclude that TW~Hya is in a phase of 
rapid spin-up as its large-scale dipole field progressively vanishes.  
\end{abstract}

\begin{keywords} 
stars: magnetic fields --  
stars: formation -- 
stars: imaging -- 
stars: rotation -- 
stars: individual:  TW~Hya --
techniques: polarimetric
\end{keywords}

\section{Introduction} 
\label{sec:int}

Magnetic fields play a significant role throughout the life of stars, all the way 
from the cradle to the grave \citep[e.g.,][for a review]{Donati09}.  
The impact of magnetic fields is strongest during the very first stages of their life, 
when protostars and their surrounding planetary systems form out of the collapse of 
giant molecular clouds;  in particular, fields are presumably efficient at slowing 
down the cloud collapse, at inhibiting the subsequent fragmentation and at dissipating 
the cloud angular momentum through magnetic braking and the magnetised outflows and 
collimated jets generated as part of the formation process \citep[e.g.,][]{Andre09}.  
In the latter phases, when low-mass protostars are still actively collecting mass 
from their surrounding accretion disc, the large-scale magnetic fields that the 
protostars generate (through dynamo processes) are strong enough to disrupt the 
inner disc regions, funnel the disc material to the stellar surface and drastically 
brake the rotation of the protostar \citep[see, e.g.,][for a review]{Bouvier07}.  

Magnetic fields at the surfaces of cTTSs were first reported indirectly, through 
the detection of various spectral proxies, usually continuum or line 
emission throughout the whole electromagnetic spectrum, from X-rays to radio 
wavelengths.  It is only about 2 decades ago that fields of cTTSs were first detected 
directly, i.e., through the Zeeman broadening of spectral lines, and found to reach 
typical magnetic intensities of several kG \citep[e.g.,][for an overview]{Johns07}.  
However, the actual large-scale topologies of these fields, controlling in particular how 
the fields couple the protostars to their discs, redirect the accreted disc material 
and subsequently brake the stellar rotation, remained unclear until recently.  
Thanks to the advent of sensitive high-resolution spectropolarimeters dedicated to 
the study of stellar magnetic fields, it is now possible to monitor polarised Zeeman 
signatures of cTTSs in various spectral lines (e.g., those probing accreting and 
non-accreting regions at the stellar surface) and use such time-series to reconstruct 
the parent large-scale magnetic topologies of cTTSs  (following the principles of 
tomographic imaging);  this option offers in particular a more quantitative way of 
studying magnetospheric accretion processes of cTTSs.  

This is the precisely the main goal of the international MaPP (Magnetic Protostars and 
Planets) project, designed to investigate (through a first survey of a dozen targets) 
how the large-scale magnetic topologies of cTTSs vary with stellar parameters such as 
mass, age, rotation and accretion rates \citep[e.g.,][]{Donati10b, Donati11}.  
MaPP also aims at providing an improved theoretical description of the observational 
results, using both analytical modelling and numerical simulations \citep[see, 
e.g.,][]{Gregory10, Romanova11} -- the ultimate ambition being to obtain 
a deeper understanding of magnetic field generation and magnetospheric accretion processes 
in cTTSs and of their impact on the formation of low-mass stars.  Results up to now 
demonstrate in particular that the large-scale magnetic fields of cTTSs closely resemble 
those of low-mass main sequence dwarfs (once both samples are matched according to their 
internal structure), with high-mass fully-convective cTTSs (with masses ranging from about 
0.5 to 1.0~\msun\ typically, for ages of a few Myr) hosting strong axisymmetric dipoles, 
while partly-convective cTTSs (of either higher masses and/or ages) exhibit a more complex 
field and a much weaker (potentially non-axisymmetric) dipole component.  

The new study presented in this paper focusses on the evolved cTTS TW~Hya, 
whose relative proximity to the Earth and intense magnetic fields \citep{Yang05, Yang07} 
make it an obvious target of investigation for MaPP.  With an age of about 8~Myr 
\citep[e.g.,][]{Torres08} and a mass of about 0.8~\msun\ 
(see Sec.~\ref{sec:tw}), TW~Hya can be viewed as an evolved version of 2 other prototypical 
cTTSs, namely AA~Tau and BP~Tau, and can thus serve as a landmark to test new generation 
models of stellar formation including the effect of magnetic fields.  
After summarising the main characteristics of TW~Hya, 
(Sec.~\ref{sec:tw}), we describe the new spectropolarimetric observations we 
collected (Sec.~\ref{sec:obs}) and outline the associated temporal variability 
and rotational modulation (Sec.~\ref{sec:var}).  We then detail the modelling of these 
data with our magnetic imaging code (Sec.~\ref{sec:mod}), briefly compare our 
results with those published already and conclude with a 
updated summary of what MaPP results tell us about how magnetic fields impact the 
formation of Sun-like stars (Sec.~\ref{sec:dis}).

\section{TW~Hya }
\label{sec:tw}

TW~Hya (K6V) is one of the low-mass cTTSs located nearest to the Earth 
(at a distance of $57\pm7$~pc);  it belongs to a loose association of young stars 
after which it was named (the TW~Hya association), whose age is estimated to be 
$\simeq$8~Myr, i.e., somewhat younger than the $\beta$~Pic moving group, another nearby 
young association \citep[e.g.,][]{Torres08}. 

Despite being older than most prototypical cTTSs, TW~Hya is still surrounded by 
a relatively massive accretion disc visible at IR and radio wavelengths, and subject 
to intense scrutiny in the last decade \citep[e.g.,][]{Qi04, Thi10}.  With a typical 
mass of a few Jupiter masses, this disc is found to be inclined at $\simeq$10\degr\ 
to the plane of the sky;  it consists at least of an optically thin inner cavity extending 
outwards up to about 4~AUs and of an optically thick outer disc extending to distances 
of up to about 200~AUs \citep[e.g.,][]{Calvet02, Akeson11}, suggesting that planet formation 
is ongoing.  The gas-to-dust mass in the disc may be lower than the standard interstellar 
value, suggesting that a significant fraction of the primordial disc gas has been 
evaporated due to the strong X-ray flux from the central protostar \citep{Thi10}.  

The photospheric temperature of TW~Hya estimated from the strength of spectral lines is 
found to be around 4150~K \citep{Torres03, Yang05}, only slightly hotter than the value of 
4000~K that the $V-R$ and $V-I$ photometric colors indicates \citep{Rucinski83, Rucinski08};  we 
therefore choose a temperature of $4075\pm75$~K as a compromise.  From this temperature, 
evolutionary models \citep[e.g.,][]{Siess00} indicate that the parent protostar 
(whose contraction is still mostly isothermal at an age of 8--10~Myr) has a typical mass 
of 0.8~\msun, a logarithmic luminosity (with respect to the Sun) of --0.45 and radius of 
1.1~\rsun.  This agrees well with the brightest photometric level recorded for TW~Hya 
\citep[$\mV=10.6$, e.g.,][]{Rucinski08}, the distance modulus derived from the Hipparcos 
distance (equal $3.75\pm0.25$) and the bolometric correction corresponding to the 
photospheric temperature \citep[$-0.95\pm0.05$, e.g.,][]{Bessell98}, yielding a bolometric 
magnitude of $5.90\pm0.25$, i.e., a logarithmic luminosity (with respect to the 
Sun) of $-0.46\pm0.10$ (see Fig.~\ref{fig:hrd}).  Evolutionary models also indicate that, at this age, 
TW~Hya is likely no longer fully-convective and hosts a radiative core extending up to 
$\simeq$0.35~\rstar\ in radius \citep{Siess00}.  

\begin{figure}
\includegraphics[scale=0.35,angle=-90]{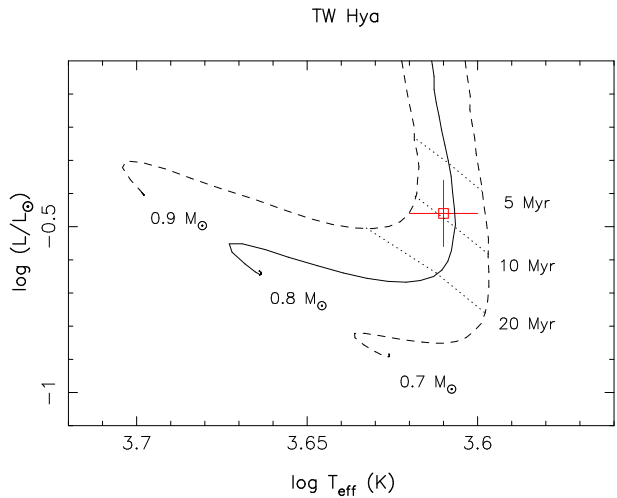}
\caption[]{Observed (open square and error bars) location of TW~Hya in the HR diagram.  
The PMS evolutionary tracks and corresponding isochrones \citep{Siess00} assume solar 
metallicity and include convective overshooting.  } 
\label{fig:hrd}
\end{figure}

The spectrum of TW~Hya was recently found to undergo regular radial velocity (RV) 
fluctuations with a period of 3.56~d, first attributed to a close-in giant planet 
\citep{Setiawan08} then to the presence of dark spots at the surface of the star 
\citep[better reproducing in particular how the RV amplitude varies with 
wavelength,][]{Huelamo08}.  It therefore strongly suggests that the period of the 
RV fluctuations is the rotation period of TW~Hya, implying an equatorial rotation 
velocity of 15.6~\kms\ (given the assumed radius of 1.1~\rsun, see above).  
Numerous estimates of the line-of-sight projected equatorial rotation velocity 
(noted \vsini\ where $i$ is the inclination of the rotation axis to the line of sight) 
are available in the literature, ranging from 4 to 15~\kms\ \citep{Torres03};  
the most reliable of these estimates (those including a basic modelling of 
the micro- and/or macro-turbulent broadening of the spectral lines in particular) 
are all on the low side of this range \citep[between 4 and 6~\kms, e.g.,][]{Alencar02, 
Torres03, Yang05}, consistent with the fact that TW~Hya is seen mostly pole-on (as the 
disc itself).  Our own estimate ($4\pm1$~\kms, see Sec.~\ref{sec:mod}) implies 
$i\simeq15\degr$, in agreement with previous estimates and with the inclination of 
the disc \citep[e.g.,][]{Alencar02}.  

From both magnetic broadening of spectral lines and Zeeman signatures of emission 
lines used as accretion proxies, kG magnetic fields were recently reported at the 
surface of TW~Hya \citep{Yang05, Yang07}, making it an obvious target for studies of 
magnetospheric accretion processes. 
High-resolution X-ray spectra of TW~Hya exhibit numerous spectral lines formed at low 
and high temperatures and allowing one to probe both the accretion shock and the corona 
\citep[e.g.,][]{Kastner02, Brickhouse10};  the low-temperature component of the X-ray spectrum 
(related to the accretion process) is particularly prominent on TW~Hya (with respect 
to other cTTSs) making it an obvious laboratory for studying the physics of magnetospheric 
accretion processes.  Observations are found to agree reasonably well with the 
predictions of theoretical models \citep[e.g.,][]{Gunther07b}.  

From the equivalent widths and the corresponding line fluxes of the optical emission lines 
(taken as accretion proxies) and using the published empirical correlations 
between lines and accretion fluxes \citep{Fang09}, we can derive an estimate of the 
logarithmic mass accretion rate (in units of \mspy) at the surface of TW~Hya, that we find 
to be equal to $-8.9\pm0.4$ at both epochs (see Sec.~\ref{sec:var}), in agreement with 
independent estimates from optical proxies \citep{Curran11}.  As usual, this is larger than 
the estimate derived from X-ray data \citep[equal to $-9.74\pm0.05$,][]{Curran11}.  

Optical veiling, i.e., the apparent weakening of the photospheric spectrum (presumably 
caused by accretion), is significant in TW~Hya and typically ranges from 30 to 100\% during 
both observing runs, in agreement with previous published reports \citep[e.g.,][]{Yang05}.

\section{Observations}
\label{sec:obs}

Spectropolarimetric observations of TW~Hya were collected at two different epochs, first from 
2008 March 15 to 28, then from 2010 February 23 to March 08, using the high resolution 
spectropolarimeter ESPaDOnS on the Canada-France Hawaii Telescope (CFHT).  ESPaDOnS collects 
stellar spectra spanning the entire optical domain (from 370 to 1,000~nm) at a resolving power 
of 65,000 (i.e., 4.6~\kms) and with a spectral sampling of 2.6~\kms, in either circular or linear 
polarisation \citep{Donati03}.
In 2008, a total of 13 circular polarisation spectra were collected over a timespan of 14 nights, 
but with fairly irregular time sampling;  in 2010, another set of 13 circular polarisation spectra 
were collected, again over 14 nights, but this time with at a much more regular rate (of about 1 
spectrum per night).  
All polarisation spectra (except the first 3 in 2008 March, affected by a setup problem) 
consist of 4 individual subexposures (each 
lasting 815~s and 840~s in 2008 and 2010 respectively) taken in different polarimeter configurations 
to allow the removal of all spurious polarisation signatures at first order.

All raw frames are processed with {\sc Libre~ESpRIT}, a fully automatic reduction
package/pipeline available at CFHT.  It automatically performs optimal
extraction of ESPaDOnS unpolarized (Stokes $I$) and circularly polarized (Stokes $V$)
spectra grossly following the procedure described in \citet{Donati97b}.
The velocity step corresponding to CCD pixels is about 2.6~\kms;  however, thanks
to the fact that the spectrograph slit is tilted with respect to the CCD lines,
spectra corresponding to different CCD columns across each order feature a
different pixel sampling.  {\sc Libre~ESpRIT} uses this opportunity to carry out
optimal extraction of each spectrum on a sampling grid denser than the original
CCD sampling, with a spectral velocity step set to about 0.7 CCD pixel
(i.e.\ 1.8~\kms).
All spectra are automatically corrected of spectral shifts resulting from
instrumental effects (e.g., mechanical flexures, temperature or pressure variations)
using telluric lines as a reference.  Though not perfect, this procedure provides
spectra with a relative RV precision of better than 0.030~\kms\
\citep[e.g.,][]{Donati08b}.

\begin{table}
\caption[]{Journal of observations collected in 2008 March and 2010 March.  
Each observation consists of a sequence of 4 subexposures (each lasting 815~s 
and 840~s in 2008 and 2010 respectively), except for the first 3 in 2008 March for which only 
2 subexposures (of 815~s each) could be used (following an instrument setup problem).  
Columns $1-4$ respectively list the UT date, the Heliocentric Julian Date (HJD) and 
UT time (both at mid-exposure), and the peak signal to noise ratio (per 2.6~\kms\ 
velocity bin) of each observation.  
Column 5 lists the rms noise level (relative to the unpolarized continuum level 
$I_{\rm c}$ and per 1.8~\kms\ velocity bin) in the circular polarization profile 
produced by Least-Squares Deconvolution (LSD), while column~6 indicates the 
orbital/rotational cycle associated with each exposure (using the ephemeris given by 
Eq.~\ref{eq:eph}).  }   
\begin{tabular}{cccccc}
\hline
Date & HJD          & UT      &  \sn\  & $\sigma_{\rm LSD}$ & Cycle \\
2008 & (2,454,500+) & (h:m:s) &      &   (\ptt)  & (0+) \\
\hline
Mar 15 & 40.97830 & 11:20:12 & 100 & 6.4 & 0.274 \\
Mar 16 & 41.84633 & 08:10:10 & 150 & 3.4 & 0.517 \\
Mar 16 & 41.94681 & 10:34:52 & 170 & 2.9 & 0.546 \\
Mar 20 & 45.90219 & 09:30:41 & 250 & 2.1 & 1.654 \\
Mar 20 & 45.95210 & 10:42:34 & 250 & 2.1 & 1.668 \\
Mar 23 & 48.82940 & 07:45:57 & 240 & 2.2 & 2.474 \\
Mar 23 & 48.97383 & 11:13:56 & 260 & 2.1 & 2.515 \\
Mar 26 & 51.86791 & 08:41:31 & 250 & 2.2 & 3.326 \\
Mar 27 & 52.82173 & 07:35:04 & 260 & 1.9 & 3.593 \\
Mar 27 & 52.96303 & 10:58:32 & 240 & 2.2 & 3.633 \\
Mar 28 & 53.81423 & 07:24:18 & 250 & 2.0 & 3.871 \\
Mar 28 & 53.90863 & 09:40:14 & 170 & 3.5 & 3.898 \\
Mar 28 & 53.95311 & 10:44:17 & 180 & 3.3 & 3.910 \\
\hline
Date & HJD          & UT      &  \sn\  & $\sigma_{\rm LSD}$ & Cycle \\
2010  & (2,455,200+) & (h:m:s) &      &   (\ptt)  & (199+) \\
\hline
Feb 23 & 51.00057 & 11:52:33 & 240 & 2.2 & 0.255 \\
Feb 24 & 52.04170 & 12:51:43 & 260 & 2.1 & 0.546 \\
Feb 25 & 52.96333 & 10:58:51 & 250 & 2.4 & 0.805 \\
Feb 26 & 54.01997 & 12:20:22 & 260 & 2.3 & 1.101 \\
Feb 27 & 54.93640 & 10:19:60 & 240 & 2.4 & 1.358 \\
Feb 28 & 56.03571 & 12:42:58 & 240 & 2.2 & 1.666 \\
Mar 01 & 57.00578 & 11:59:51 & 220 & 2.5 & 1.938 \\
Mar 02 & 58.01739 & 12:16:33 & 170 & 3.7 & 2.221 \\
Mar 03 & 58.95437 & 10:45:47 & 160 & 3.7 & 2.484 \\
Mar 04 & 60.01111 & 12:07:28 & 220 & 2.5 & 2.780 \\
Mar 05 & 60.95526 & 10:47:01 & 190 & 3.2 & 3.044 \\
Mar 07 & 62.98278 & 11:26:38 & 220 & 2.6 & 3.613 \\
Mar 08 & 63.95641 & 10:48:39 & 160 & 3.8 & 3.885 \\
\hline
\end{tabular}
\label{tab:logesp}
\end{table}

The peak signal-to-noise ratios (\sn, per 2.6~\kms\ velocity bin) achieved on the
collected spectra range between 100 and 260 depending on weather/seeing conditions 
(and on the instrumental setup), with a median value of about 240.  
The full journal of observations is presented in Table~\ref{tab:logesp}.  

Following \citet{Huelamo08}, we assume that the period of the RV variations is the 
rotation period.  Rotational cycles $E$ are computed from Heliocentric Julian Dates (HJDs) 
according to the ephemeris \citep[note the different initial HJD than that used 
by][]{Huelamo08}:  
\begin{equation}
\mbox{HJD} = 2454540.0 + 3.5683 E.  
\label{eq:eph}
\end{equation}
Coverage of the rotation cycle is only moderate in 2008 March (with a phase gap of 
about 1/3 of the cycle) and good in 2010 March (see Table~\ref{tab:logesp}).  
As a result of the difference in the assumed initial HJD, our rotation cycles are 
ahead of those of \citet{Huelamo08} by 2.774 cycles (at both epochs);  
maximum RVs should thus occur around phase 0.1 in our ephemeris.  

\begin{figure}
\includegraphics[scale=0.35,angle=-90]{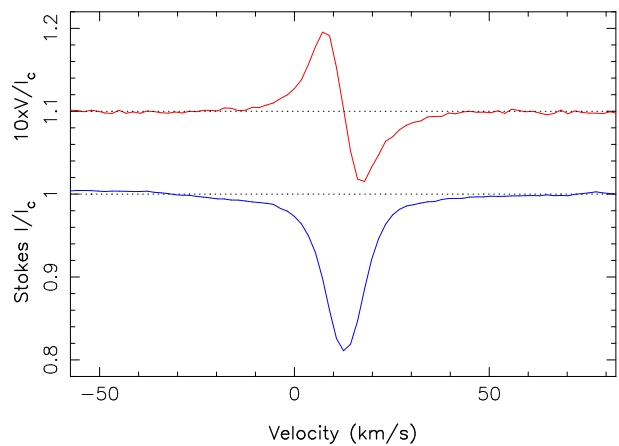}
\caption[]{LSD circularly-polarized (Stokes $V$) and unpolarized (Stokes $I$)
profiles of TW~Hya (top/red, bottom/blue curves respectively) collected on 2010 
February 26 (cycle 199+1.101).  A strong Zeeman signature (with a full 
amplitude of 1.8\%) is detected in the LSD 
Stokes $V$ profile, in conjunction with the unpolarised line profile.  
The mean polarization profile is expanded by a factor of 10 and shifted upwards
by 1.1 for display purposes.  }
\label{fig:lsd}
\end{figure}

Least-Squares Deconvolution \citep[LSD,][]{Donati97b} was applied to all
observations.   The line list we employed for LSD is computed from an {\sc
Atlas9} LTE model atmosphere \citep{Kurucz93} and corresponds to a K7V 
spectral type ($\teff=4,000$~K and  $\logg=4.5$) appropriate for TW~Hya.  
Only moderate to strong atomic spectral lines (with line-to-continuum core 
depressions larger than 40\% prior to all non-thermal broadening) are included 
in this list;  spectral regions with strong lines mostly formed outside the 
photosphere (e.g., Balmer, He, \caii\ H, K and IRT lines) and/or heavily crowded 
with telluric lines were discarded.  
Altogether, about 6,500 spectral features (with about 40\% from \fei) are used 
in this process.  
Expressed in units of the unpolarized continuum level $I_{\rm c}$, the average 
noise levels of the resulting Stokes $V$ LSD signatures range from 1.9 to 
6.4$\times10^{-4}$ per 1.8~\kms\ velocity bin (median value 2.4$\times10^{-4}$).  
Zeeman signatures are detected at all times in LSD profiles (see Fig.~\ref{fig:lsd} 
for an example) and in most accretion proxies (see Sec.~\ref{sec:var}).  

\begin{figure*}
\center{\includegraphics[scale=0.80,angle=-90]{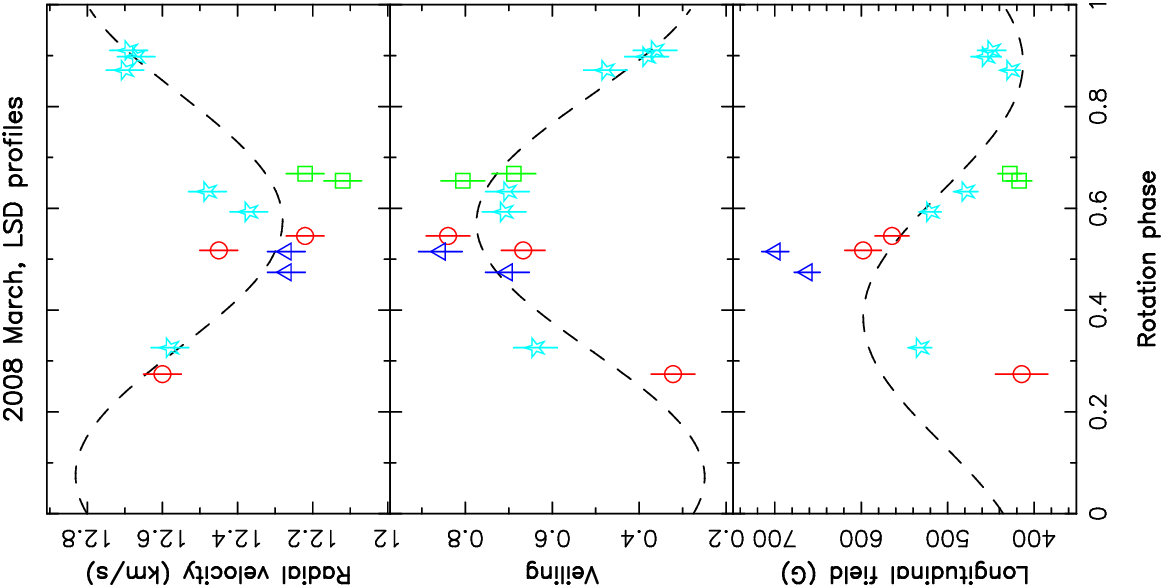}\hspace{4mm}
\includegraphics[scale=0.80,angle=-90]{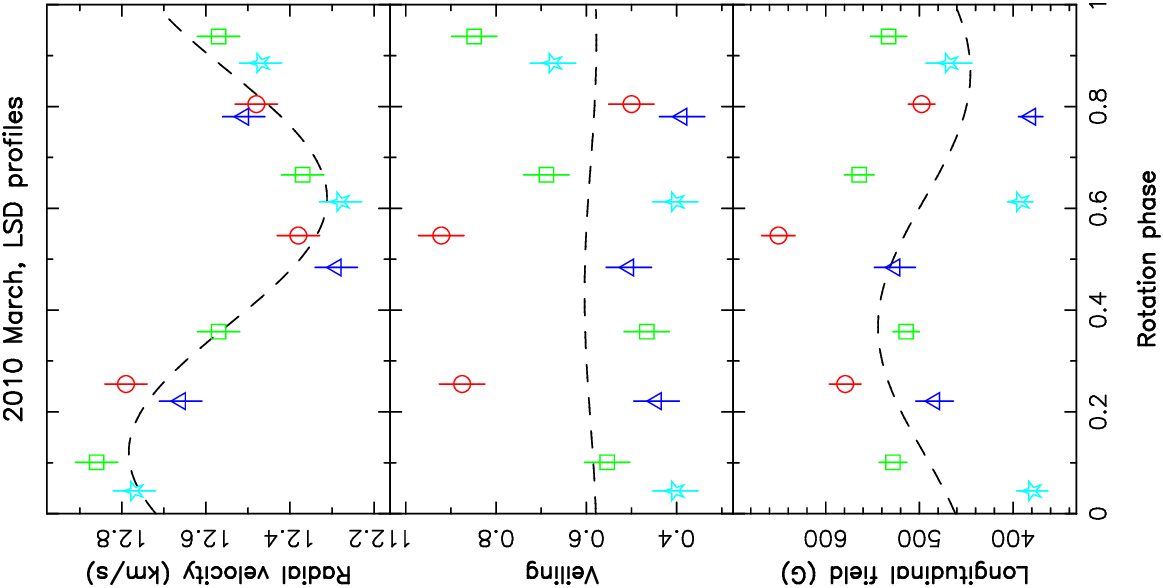}}
\caption[]{Rotational modulation of the RV (top row), veiling (second row) 
and longitudinal field (bottom row) derived from the LSD photospheric profiles 
of TW~Hya in 2008 March (left panels) and in 2010 March (right panels).  
Data collected during rotational cycles 0, 1, 2 and 3 are respectively shown 
with red circles, green squares, dark-blue triangles and light-blue stars.  
Fits with sine/cosine waves are included (and shown as dashed lines) to outline 
(whenever significant) the amount of variability attributable to rotational modulation.  
$\pm1$~$\sigma$ error bars on data points are also shown.  }
\label{fig:var1}
\end{figure*}

\begin{figure*}
\center{\includegraphics[scale=0.80,angle=-90]{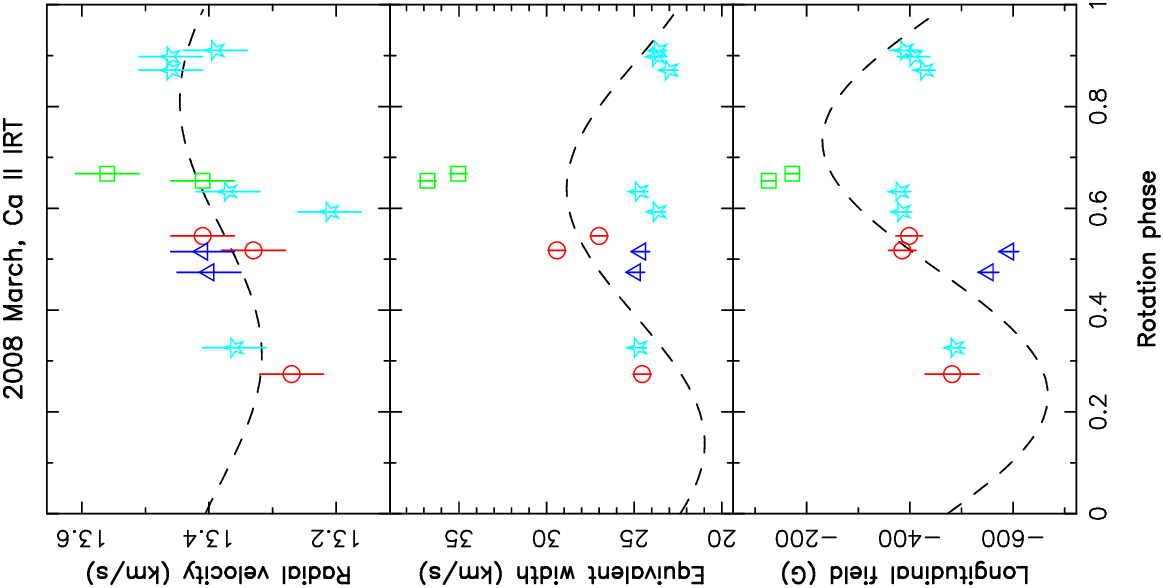}\hspace{4mm}
\includegraphics[scale=0.80,angle=-90]{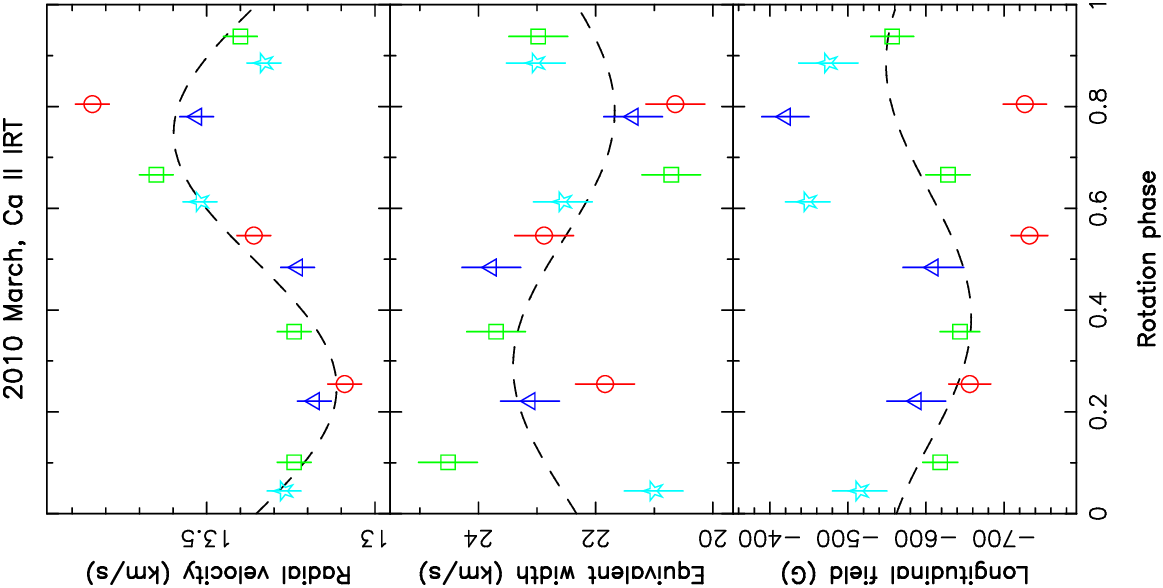}}
\caption[]{Same as Fig.~\ref{fig:var1} for the \caii\ IRT LSD profiles. } 
\label{fig:var2}
\end{figure*}

\section{Spectroscopic variability}
\label{sec:var}

In this section, we present a basic description of how the spectral lines of TW~Hya, 
and in particular the photospheric LSD profiles and the selected accretion proxies 
(i.e., the \caii\ infrared triplet or IRT, the \hei\ $D_3$ line and the first 2 
lines of the Balmer series), vary with time and with rotation phase.  The idea is 
to look at how the equivalent widths, the RVs and the average magnetic fluxes 
of the various profiles are modulated with rotation, to get a rough, intuitive 
understanding of how the large-scale field is structured and oriented, and how 
cool photospheric spots and hot chromospheric accretion regions are distributed over 
the stellar surface - to be compared with the results of the full tomographic 
imaging analysis carried out in the following section.  

\subsection{LSD photospheric profiles}

As shown on Fig.~\ref{fig:var1} (top and middle panels), the photospheric 
Stokes $I$ LSD profiles exhibit significant variability at the two observing epochs, 
both in position (RV) and strength (as a result of veiling)\footnote{We assume a 
conservative error bar of $\pm0.05$~\kms\ and $\pm2$\% on the relative RV and the 
relative strength of photospheric LSD and \caii\ IRT profiles;  in particular, the 
error bar on the RVs of LSD profiles is larger than the formal error bar of 0.03~\kms, 
but comparable to the typical RV dispersion quoted by \citet{Huelamo08}.  }.  
Showing no more than a 
moderate level of intrinsic variability, RVs are clearly modulated by rotation, varying 
by $\simeq$0.5~\kms\ peak to peak around a mean of $12.55\pm0.10$~\kms\ and reaching a 
maximum at phase $\simeq$0.1 (corresponding to phase 0.88 in the ephemeris of 
\citealt{Huelamo08}).  This is fully compatible with the results of \citet{Huelamo08}, 
indicating that the RV variability we report is of the same origin and nature 
than that discussed by \citet{Huelamo08}, i.e., caused by a cool photospheric spot on 
TW~Hya located at high latitudes and at phase $\simeq$0.35 (0.25 cycle after RV maximum);  
moreover, it suggests that this spot is apparently persistent on timescales of at least 3~yrs.  

LSD profiles of TW~Hya are also significantly shallower than those expected from a standard 
star of similar spectral type, indicating that the spectrum is significantly veiled (as a 
likely result of accretion), at levels varying from 30\% to 100\% (with an average of about 
60\%).  This is again typical to previously published reports \citep[e.g.,][]{Yang05}.  
We find that veiling can be mainly modulated by rotation at times, as in 2008 March where 
maximum veiling is observed to occur at phase 0.6 (i.e., 0.25 cycle later than the phase 
at which the cool spot is best visible);  at other times, veiling is dominated mostly by 
intrinsic variability, as in 2010 March where no clear trend with rotation phase is 
detected (see middle panels of Fig.~\ref{fig:var1}).  

As mentioned already, Zeeman signatures are detected at all times in Stokes $V$ LSD 
profiles of TW~Hya, showing a fairly canonical shape (i.e., antisymmetric with respect to 
the line centre) and an unusually large amplitude of almost 2\% peak to peak (see, e.g., 
Fig.~\ref{fig:lsd}).  
The line-of-sight projected component of the field averaged over the visible stellar 
hemisphere and weighted by brightness inhomogeneities \citep[called the longitudinal 
field and estimated from the first moment of the Stokes $V$ profile, e.g.,][]{Donati97b} 
ranges from 380 to 700~G, with typical error bars of $\simeq$15~G.  
Note that our longitudinal field estimates are significantly larger than those found 
in previously published studies \citep[reporting values $\leq$150~G,][]{Yang07};  
this discrepancy can potentially be attributable to long-term evolution of the large-scale 
magnetic topology \citep[as suggested for a similar mismatch in the case of the cTTS 
BP~Tau,][]{Donati08}, as the data of \citet{Yang07} were collected about a decade 
earlier than ours (in 1999).  

As shown on Fig.~\ref{fig:var1} (bottom panels), the longitudinal field curves at 
both epochs include a significant amount of intrinsic variability (especially in 
2010 March), but nevertheless suggest a $\pm$10--20\% level of rotational modulation 
about the mean value (of $\simeq$500~G) with maximum longitudinal field occurring 
at phase $\simeq$0.35, i.e., when the cool spot is best visible.  Most likely, this 
intrinsic variability of the longitudinal field does not reflect a true change in 
the large-scale magnetic topology, but rather relates to modifications in the line 
formation region and related properties (e.g., veiling, cool spot location and size).  

\subsection{\caii\ IRT emission}

As in previous papers, we use core emission in \caii\ IRT lines as our main proxy of 
surface accretion even though it usually features a significant contribution from the 
non-accreting chromosphere;  previous work \citep[e.g.,][]{Donati10b, Donati11} demonstrated 
in particular that rotational modulation in Stokes $I$ and $V$ \caii\ IRT profiles of cTTSs 
correlates well with that derived from the more conventional \hei\ $D_3$ accretion 
proxy (see below), even at fairly low mass-accretion rates.  
To recover the emission component of the \caii\ IRT lines, we start by constructing a LSD-like 
weighted average of the 3 IRT lines;  we then subtract the underlying (much wider) photospheric 
absorption profiles, with a single Lorentzian fit to the far line wings \citep[see][for an 
illustration]{Donati11b}.  

The average \caii\ emission profiles are centred at a RV of $\simeq$13.40~\kms, i.e., slightly 
red-shifted (by 0.85~\kms) with respect to the photospheric LSD profiles (see Fig.~\ref{fig:var2}, 
top panels);  this is usual for cTTSs and further confirms that \caii\ emission probes 
slowly-moving plasma in the post-shock regions of accretion funnels, and located close to the 
surface of the star.  
We also observe that the RVs of the \caii\ emission profiles show, in addition to a weak level 
of intrinsic variability, a small amount of rotational modulation, with an amplitude 
comparable to (in 2010 March) or smaller than (in 2008 March) that seen in the LSD photospheric 
profiles and much smaller (by at least a factor of 15) than the rotational broadening of 
spectral lines ($\vsini\simeq4$~\kms, see Sec.~\ref{sec:mod}).  It suggests in 
particular the presence of a localised region of accretion-powered emission near the pole, 
and best viewed (at both epochs) at a rotational phase of $\simeq$0.5 (i.e., at mid transit 
between minimum and maximum RVs).  This approximately coincides with the phase of maximum 
veiling in 2008 March ($\simeq$0.6, see above) and with that at which the cool photospheric 
spot (probed through LSD profiles) is best visible ($\simeq$0.35), the associated phase shifts 
(of +0.10 and --0.15 rotation cycle respectively) being not necessarily very significant at 
high latitudes and with such a high level of intrinsic variability.  

Emission strength is also varying with time (see middle panels of Fig.~\ref{fig:var2}), with 
equivalent widths typically within the range 20--30~\kms\ (0.055--0.085~nm) and featuring at 
times short-lived sporadic bursts during which emission is temporarily enhanced by as much as 
50\% (e.g., at cycle 1.654 and 1.668 in 2008 March).  Rotational modulation is moderate, 
potentially up to 30\% peak-to-peak in 2008 March with maximum emission occurring near phase 
0.6 (in conjunction with the phase of maximum veiling, see above);  in 2010 March however, 
rotational modulation is much lower (lower than 10\% peak-to-peak) and mostly hidden by 
intrinsic variability.  


\begin{figure*}
\center{
\includegraphics[scale=0.65,angle=-90]{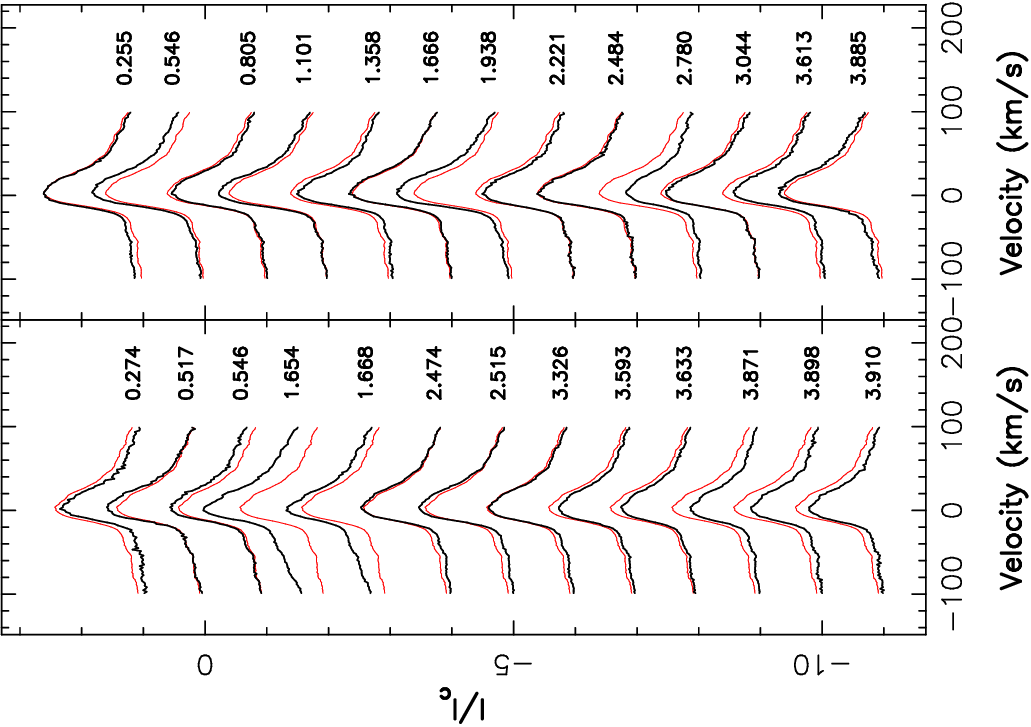}\hspace{5mm}
\includegraphics[scale=0.65,angle=-90]{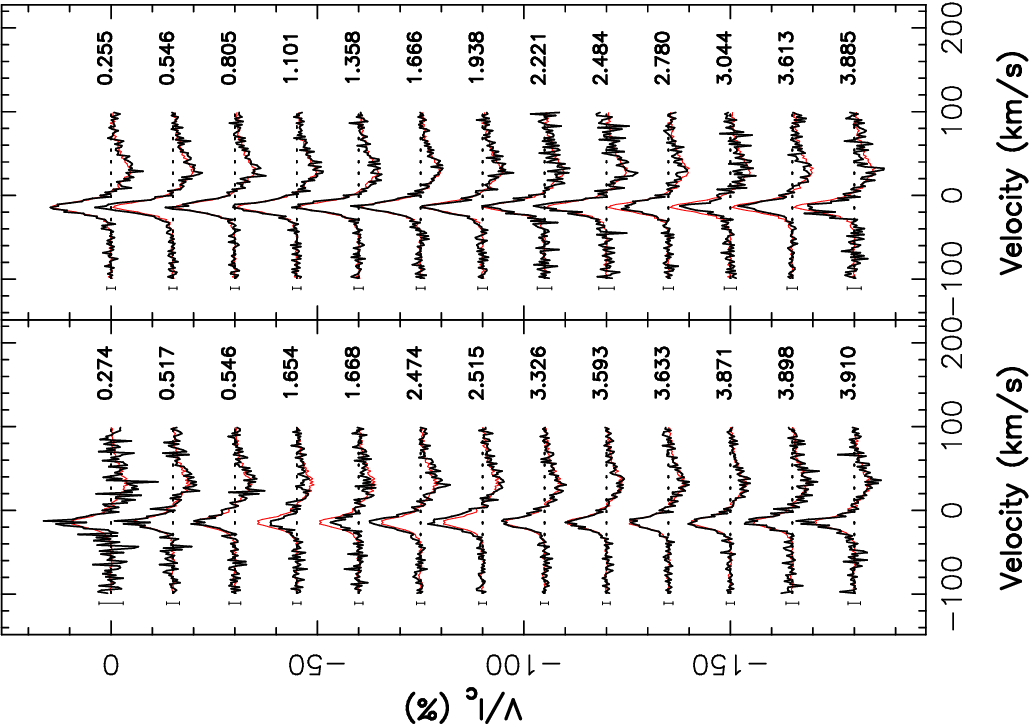}}
\caption[]{Variations of the unpolarized (Stokes $I$, left panel) and circularly-polarized 
(Stokes $V$, right panel) profiles of the \hei\ $D_3$ emission of TW~Hya in 2008 March (left 
columns of both panels) 2010 March (right columns).  The unpolarized emission profile is 
asymmetric with respect to the line centre and can be described as the sum of a central 
narrow component and a broad (often red-shifted) component (see text).  
A strong Zeeman signature (with a full amplitude of $\simeq$20\%) is detected in conjunction 
with the central narrow \hei\ emission, but not with the broad red-shifted component.
To emphasize variability, the average profile over each run is shown in red.
Rotation cycles (as listed in Table~1) and 3$\sigma$ error bars (for Stokes $V$ profiles only) 
are shown next to each profile.  }
\label{fig:he}
\end{figure*}

As for LSD profiles, clear Zeeman signatures are detected at all times in conjunction with the 
emission component of \caii\ IRT lines, with peak-to-peak amplitudes reaching up to 10\% (in 
2010 March).  The corresponding longitudinal fields range between --380 and --730~G (with typical 
error bars of about 30~G), except during the previously mentioned emission burst (when the 
longitudinal field is reduced to $-150$~G).  Note in particular the striking polarity contrast 
between the (constantly negative) longitudinal fields measured from the \caii\ emission lines 
and those (always positive) derived from the LSD profiles;  this clearly demonstrates that LSD 
profiles and IRT lines probe different areas (of different magnetic polarities) over the stellar 
surface of TW~Hya \citep[as on V2129~Oph and BP~Tau,][]{Donati11, Donati08}.  
The field values we derive from the \caii\ IRT profiles are comparable (both in strength 
and sign) to those reported in a previous paper \citep{Yang07}, in contrast to longitudinal fields 
from LSD photospheric profiles (found to be significantly stronger in our study, see above);  it 
may suggest that the potential field evolution between the two studies concerns the magnetic 
topology as a whole and not just its overall strength.  

Again, intrinsic variability is dominant at both epochs and no more than a moderate level of 
rotational modulation is observed, with strongest longitudinal fields occurring approximately
when the cool photospheric spot is best visible (i.e., around phase 0.35);  as for longitudinal 
fields estimated from LSD profiles, we suggest that this intrinsic variability reflects changes in 
the line formation region (and in particular in the accretion region) rather than in the large-scale 
magnetic topology itself.  

\begin{figure*}
\center{\includegraphics[scale=0.80,angle=-90]{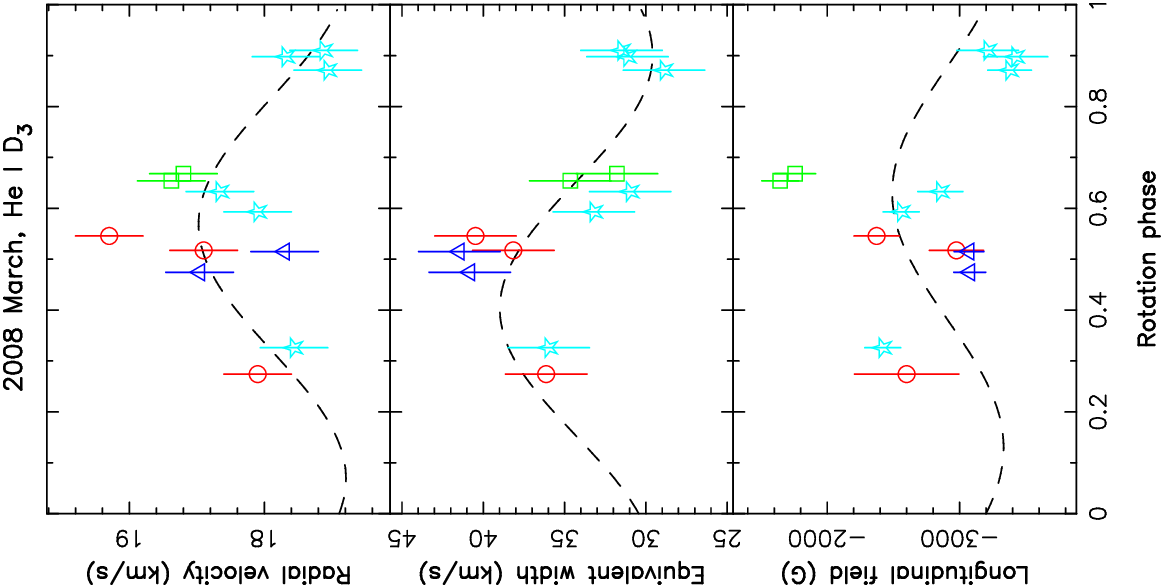}\hspace{4mm}
\includegraphics[scale=0.80,angle=-90]{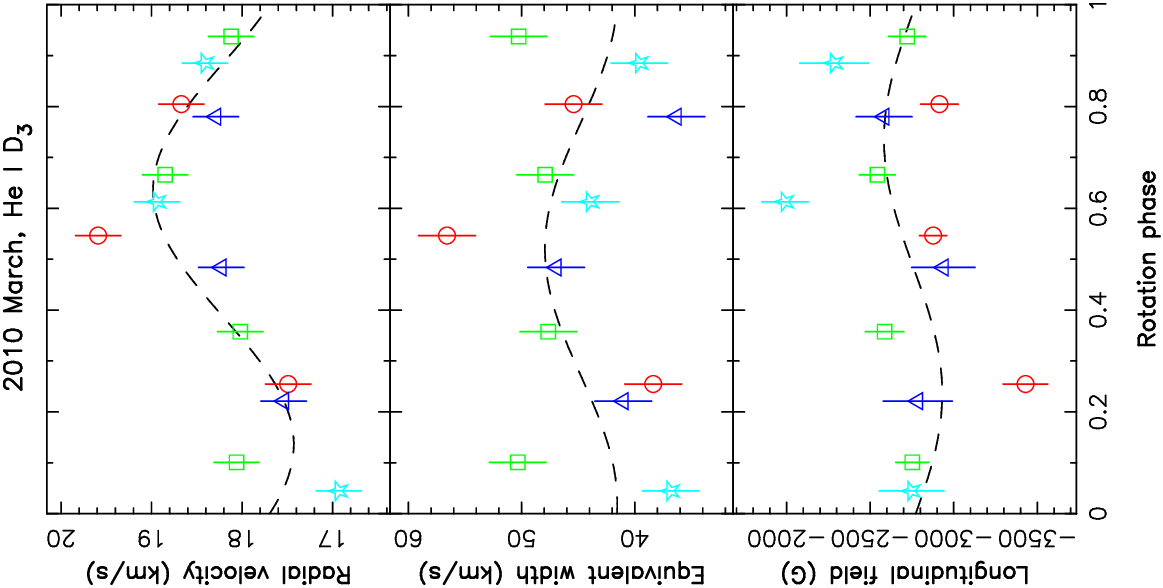}}
\caption[]{Same as Fig.~\ref{fig:var1} for the narrow component of the \hei\ $D_3$ profiles. } 
\label{fig:var3}
\end{figure*}

\subsection{\hei\ $D_3$ emission}

The \hei\ $D_3$ line, used in most cases as the reference accretion proxy, exhibits, in the 
spectrum of TW~Hya, an emission profile that is strongly asymmetric with respect to the line 
centre (see Fig.~\ref{fig:he}).  This profile can be decomposed into a narrow central emission 
centred at $\simeq$18.2~\kms\ (i.e., red-shifted by about 5.6~\kms\ with respect to the LSD 
profiles) and whose full-width at half maximum is $\simeq$36~\kms, and a broad emission component 
(full width at half maximum $\simeq$110~\kms) red-shifted by typically 20--30~\kms\ with respect 
to the narrow component;  the equivalent width of the narrow component (35 and 45~\kms\ on 
average in 2008 March and 2010 March respectively, or equivalently 0.07 and 0.09~nm) is about half 
that of the broad component ($\simeq$75~\kms\ or 0.15~nm).  Zeeman signatures with typical amplitudes 
of $\simeq$20\% are observed in conjunction with the narrow emission component, but not with the broad 
component (see Fig.~\ref{fig:he}).  Their shape strongly departs from the usual antisymmetric pattern 
(with respect to the line centre) likely indicating that the line forms in a region with significant 
velocity gradients \citep[e.g.,][]{Sanchez92}.  
We suggest that this narrow emission component comes from the postshock region in which the 
postshock plasma is strongly decelerating though still falling towards the stellar surface (hence 
the red-shifted emission and the non-antisymmetric Zeeman signature). 
Similarly, we propose that the broad emission component probes the whole accretion funnels rather than 
just the postshock region (hence the larger line width and red-shift, reflecting 
the higher plasma velocities involved).  
In the following, we focus on the narrow emission component mostly, 
being the most relevant one for our study on the large-scale field topology, photospheric spots and 
accretion regions at the surface of TW~Hya.  

The temporal variability of the narrow \hei\ emission is shown in Fig.~\ref{fig:var3} for both 
observing epochs\footnote{We assume an error bar of $\pm$0.25~\kms\ and $\pm$2.5~\kms\ on the relative 
RV and the relative strength of the narrow \hei\ emission component, i.e., 5$\times$ larger than 
those used for the \caii\ IRT profiles to reflect the reduced accuracy resulting from the 
subtraction of the broad \hei\ emission component.}.  We find that the RV of the narrow \hei\ 
emission is modulated by rotation in a similar way than the \caii\ IRT emission, suggesting 
again the presence of an accretion region located around phase 0.3--0.4 (at mid distance 
between minimum and maximum RVs), i.e., more or less cospatial with the cool photospheric spot 
probed through LSD profiles (see above).  The semi-amplitude of the RV variations reaches 
0.5--0.8~\kms\ for the narrow \hei\ emission (see Fig.~\ref{fig:var3} top panels), significantly larger than 
those seen in the \caii\ IRT lines;  this is in agreement with the much smaller dilution that the 
\hei\ line suffers from the non-accreting chromosphere.  Assuming that this dilution is negligible 
for the \hei\ line and thus that the associated RV fluctuations directly track the Doppler motion of 
the hot spot, we infer that this accretion region is off-centred with respect to the pole by 
typically 10\degr\ (for \vsini=4~\kms).  

We also find that the equivalent width of the narrow \hei\ emission 
is varying with time.  Rotational modulation is again moderate, up to about 30\% and peaking at phase 
$\simeq$0.4 in 2008 March and no larger than 10\% in 2010 March (see Fig.~\ref{fig:var3} middle
panels);  intrinsic variability is significant and dominates in 2010 March.  These conclusions 
mostly confirm what the \caii\ IRT lines already suggested (see above).  
Surprisingly, the emission burst detected in \caii\ does not seem to show up in the narrow \hei\ emission 
component, although clearly detected in the \hei\ line (broad component);  this may however reflect 
the limits of our two-component decomposition (used to describe the \hei\ line) in the 
particular case of emission bursts (where both components happen to have roughly equal RVs, see 
Fig.~\ref{fig:he}).  

The longitudinal fields we derive from the observed Zeeman signatures range from --2 to --3.5~kG 
(see Fig.~\ref{fig:var3} bottom panels) when we take into account that the narrow \hei\ 
emission is the only contributor.  Unsurprisingly, the field polarity is the same as that 
derived from \caii\ IRT emission, demonstrating that both lines are reliable probes of 
accretion regions (even though \caii\ emission suffers a strong dilution from the 
non-accreting chromosphere).  
Our \hei\ longitudinal field estimates are significantly larger than those of 
\citet{Yang07} by typically a factor of 2;  this discrepancy is however only apparent and 
reflects their assumption that the full \hei\ emission (including the broad emission, rather than 
the narrow component only) contributes to the Zeeman signatures.  

As for the \caii\ IRT emission, intrinsic variability 
dominates and only moderate rotational modulation is observed, with maximum longitudinal 
field (of about --3~kG) occurring approximately when the cool photospheric spot is best 
visible;  once again, we suspect that frequent changes in the properties of the 
accretion region are causing the enhanced dispersion of the observed longitudinal fields.  

\begin{figure*}
\center{
\includegraphics[scale=0.65,angle=-90]{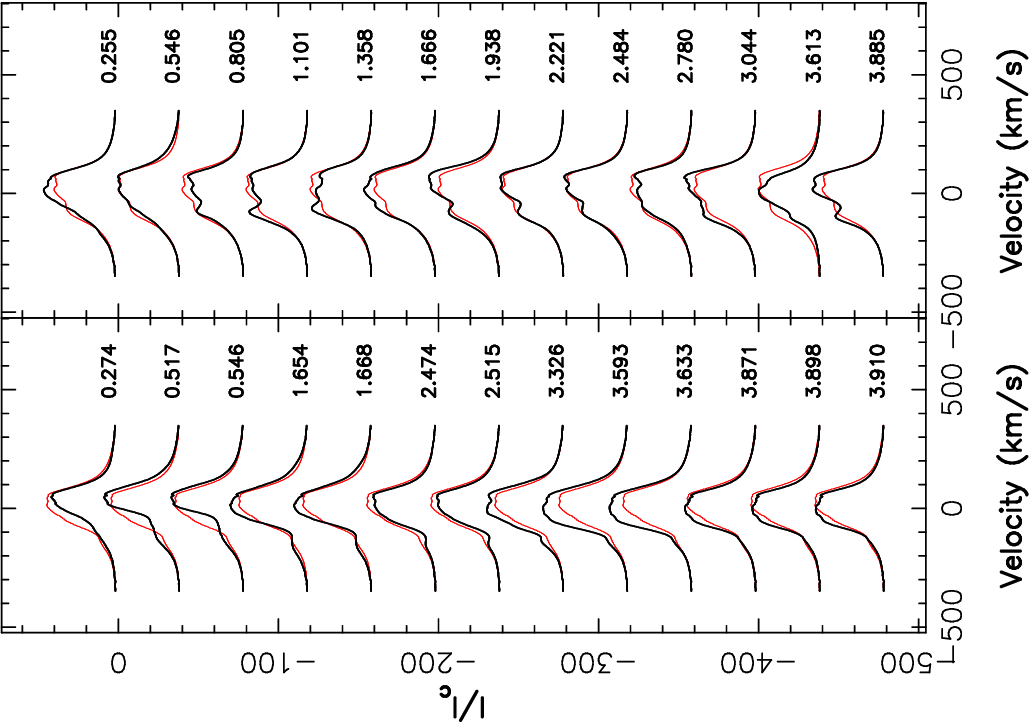}\hspace{5mm}
\includegraphics[scale=0.65,angle=-90]{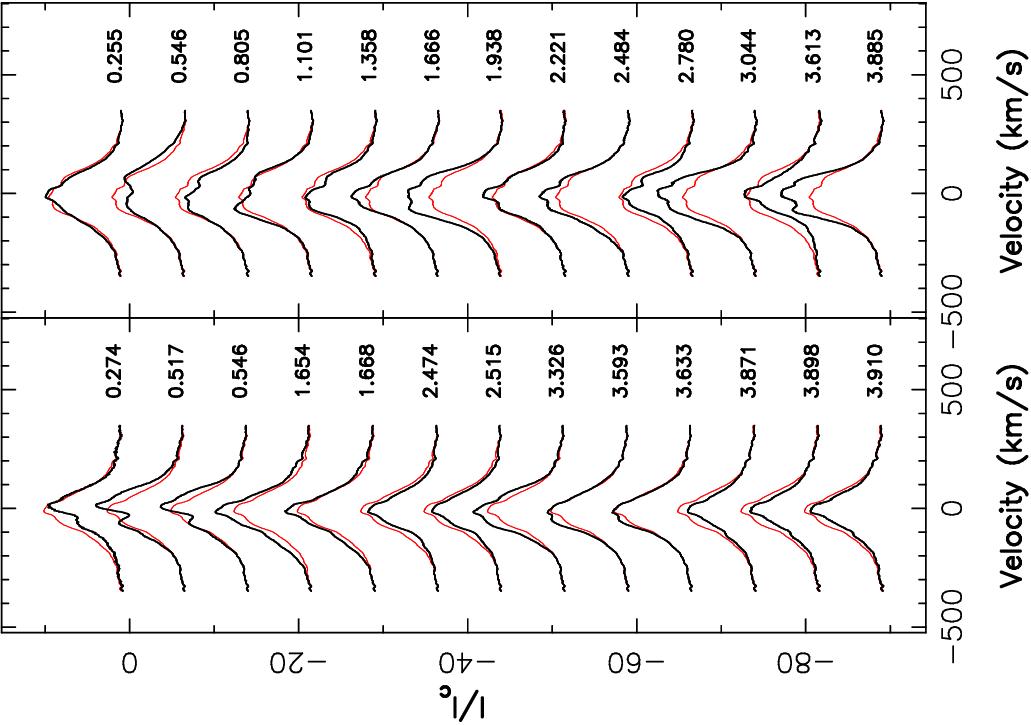}}
\caption[]{Variations of the \hal\ (left) and \hbe\ (right) lines in the spectrum of 
TW~Hya, in 2008 March (left column of both panels) and 2010 March (right column).
To emphasize variability, the average profile over each run is shown in red.
Rotation cycles (as listed in Table~1) are mentioned next to each profile.  }
\label{fig:bal}
\end{figure*}

\subsection{Balmer emission}

Balmer lines are also strongly in emission in the spectrum of TW~Hya, with \hal\ and \hbe\ 
reaching typical equivalent widths of 8,000--10,000~\kms\ (18--22~nm) and 1,700--2,100~\kms\ 
(2.8--3.4~nm) respectively.  At times, they show a conspicuous blue-shifted absorption component 
(e.g., at cycles 0.274--0.546 in 2008 March) indicating the presence of a strong wind 
escaping the star at an average projected velocity of $\simeq$50~\kms.  

Both lines exhibit significant fluctuations with time (see Fig.~\ref{fig:bal};  
however, since the profiles and emission strengths of \hal\ and \hbe\ do not convincingly 
repeat between consecutive rotation cycles (e.g., cycles 0.517 and 2.515 in 2008 March, 
cycles [199+] 0.255 and 2.221 in 2010 March), these variations cannot be attributed to 
rotational modulation and are more likely caused by intrinsic temporal evolution of the 
accretion and wind flows.  In 2008 March, the blue-shifted wind absorption is causing most 
of the observed variability in \hal\ and \hbe;  in 2010 March, the emission fluxes of both 
Balmer lines correlate well with that of the broad (and not of the narrow) \hei\ emission 
component.  

We also note that the red wing of \hbe\ is depressed at times with respect to the average profile 
(e.g., at cycle [199+] 0.805 in 2010 March).  However, these depressions never show up as true 
absorption as they regularly do on some cTTSs \citep[e.g., AA~Tau and V2129~Oph,][]{Bouvier07b, Donati11};  
we thus think it is ambiguous and potentially misleading to relate these events to the crossing of 
accretion funnels in front of the visible hemisphere of TW~Hya (as done for AA~Tau and V2129~Oph), 
as they could just as well result 
from varying projection effects of the magnetospheric funnel structure.  

\subsection{Mass-accretion rate}

From the average equivalents widths of the \caii\ IRT, \hei\ (narrow component) and \hbe\ 
lines, we can derive logarithmic line fluxes (with respect to the luminosity of the Sun \lsun), 
equal to $-4.8$, $-4.8$ and $-3.3$ respectively\footnote{To derive line fluxes from normalized 
equivalent widths, we approximate the continuum level by a Planck function at the temperature of 
the stellar photosphere. }, 
implying logarithmic accretion luminosities (with respect to \lsun) of $-2.1$, $-1.6$ and $-1.2$ 
respectively \citep[using empirical correlations from][]{Fang09}.  We thus conclude that the 
average logarithmic mass accretion rate of TW~Hya (in \mspy) is equal to $-8.9\pm0.4$ at both 
epochs, in rough agreement with the estimate of \citet{Curran11}, equal to $-9.2\pm0.2$ and also 
derived from optical lines (though from fully independent data and a different analysis).  

Mass accretion rates can in principle also be estimated (though less accurately) through the full 
width of H$\alpha$ at 10\% height \citep[e.g.,][]{Natta04, Cieza10}.  In the case of TW~Hya, 
\hal\ shows a full width of $\simeq$400~\kms\ on average, implying a logarithmic mass accretion rate            
estimate of $-9.0\pm0.6$ (in \mspy) in agreement with our previous estimate.

\section{Magnetic modelling}
\label{sec:mod}

\subsection{Overview of the method}

Now that the temporal variability of photospheric LSD and \caii\ IRT emission profiles has been 
examined in detail, we aim at converting the corresponding data sets into maps of the large-scale magnetic 
topology, as well as distributions of surface cool spots and of chromospheric accretion regions, 
at the surface of TW~Hya.  To achieve this, we apply our tomographic imaging technique, described 
extensively in previous similar studies \citep[e.g.,][]{Donati10b, Donati11}.  

Following the principles of maximum entropy, our code automatically and simultaneously derives the 
simplest magnetic topology, photospheric brightness image and accretion-powered \caii\ emission 
map compatible with the series of observed Stokes $I$ and $V$ LSD and \caii\ IRT 
profiles.  The reconstruction process is iterative and proceeds by comparing at each step the synthetic
Stokes $I$ and $V$ profiles corresponding to the current images with those of the observed data set.
The magnetic topology is described through its poloidal and toroidal components expressed
as spherical-harmonics (SH) expansions \citep[e.g.,][]{Donati06b}.  
The spatial distributions of photospheric
brightness (with respect to the quiet photosphere) and those of accretion-powered \caii\ emission (in 
excess of and with respect to that produced by the quiet chromosphere) are modelled as series of 
independent pixels (typically a few thousand) on a grid covering the visible surface of the star 
(with spots in the brightness image assumed to be darker/cooler than the quiet photosphere and 
accreting regions in the \caii\ emission maps supposed to be brighter than the quiet 
chromosphere).

Synthetic profiles are computed by summing up the elementary spectral contributions from all image
pixels over the visible stellar hemisphere, taking into account the relevant local parameters 
of the corresponding grid cells (e.g., brightness, accretion-powered excess emission, magnetic field 
strength and orientation, radial velocity, limb angle, projected area).  Since the problem is partly 
ill-posed, we stabilise the inversion process by using an entropy criterion (applied to the SH
coefficients and to the brightness/excess emission image pixels) aimed at selecting the magnetic 
topology and images with minimum information among all those compatible with the data.
The relative weights attributed to the various SH modes can be imposed, e.g., for purposely
producing antisymmetric or symmetric field topologies with respect to the centre of the star
\citep[by favouring odd or even SH modes,][]{Donati07, Donati08}.
More details concerning the specific description of local profiles used in the model can be 
found in \citet{Donati10b}.  

\begin{figure*}
\center{\hbox{
\includegraphics[scale=0.37,angle=-90]{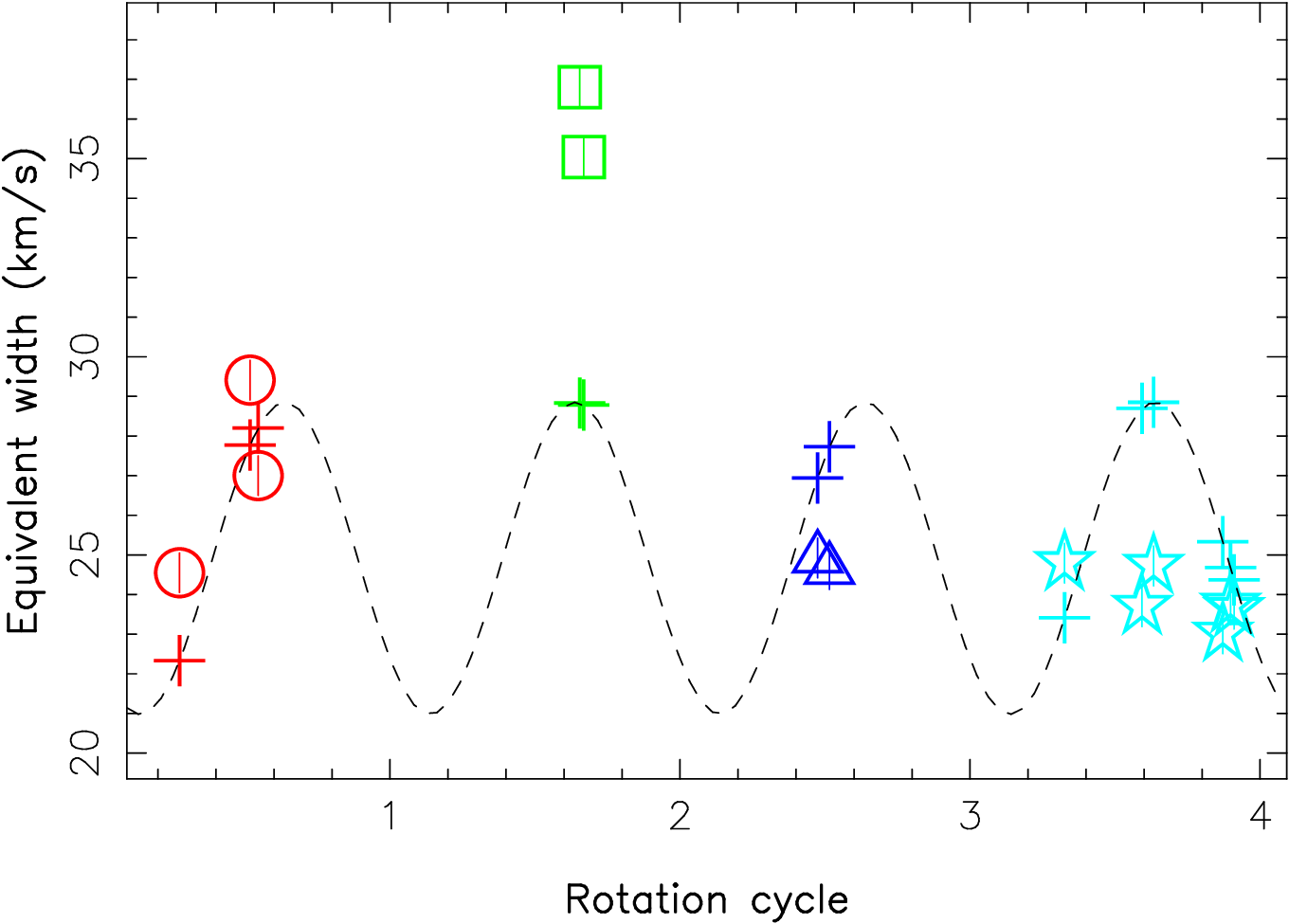}\hspace{5mm}
\includegraphics[scale=0.37,angle=-90]{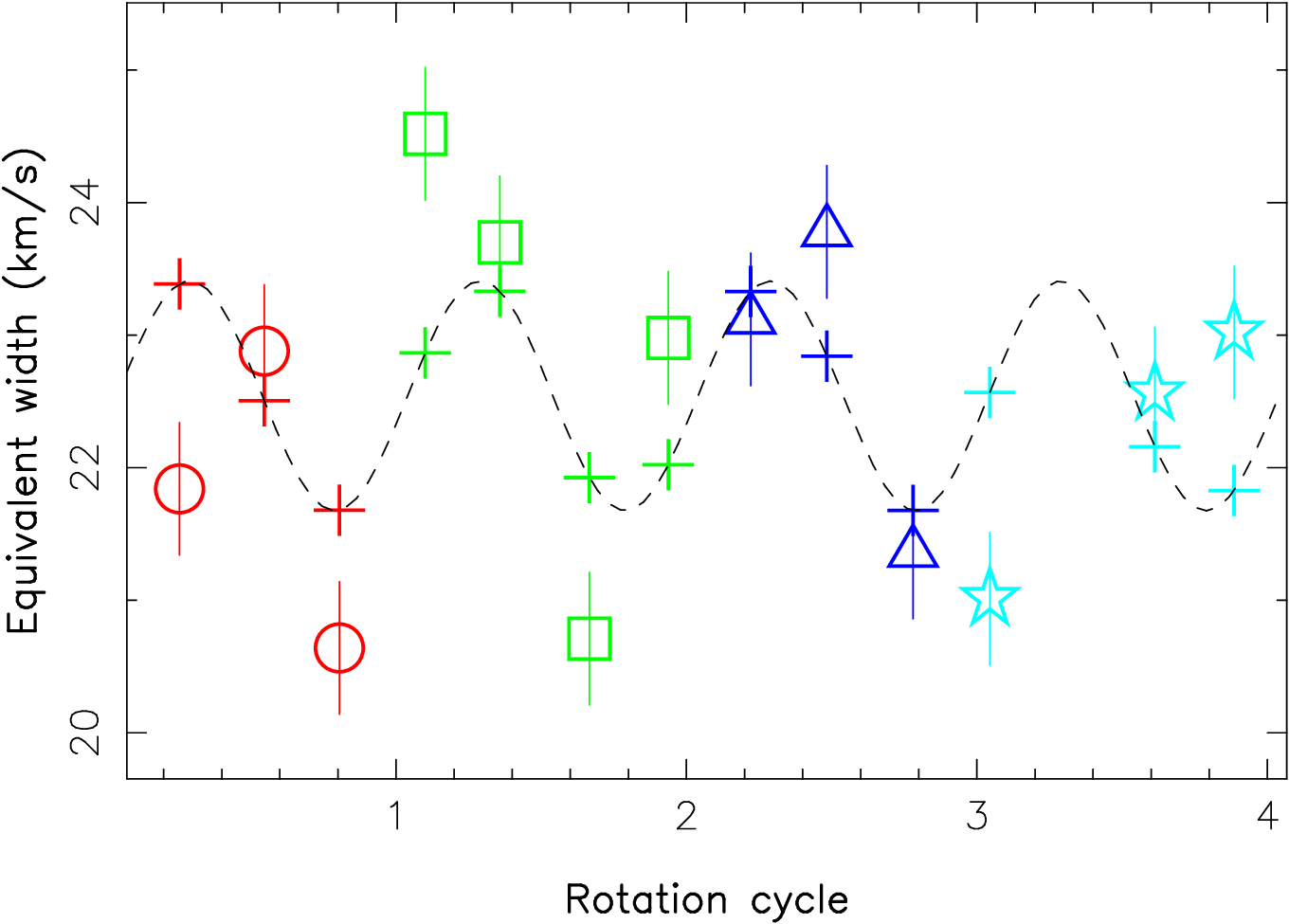}}}
\caption[]{Measured (open symbols) and fitted (pluses) equivalent widths of
the \caii\ IRT LSD profiles of TW~Hya in 2008 March (left panel) and 2010 March (right panel).
The model wave (dashed line) providing the best (sine+cosine) fit to
the data presumably traces rotational modulation, while the deviation from the fit 
illustrates the strength of intrinsic variability.  The open symbols are defined as 
described in Fig.~\ref{fig:var1}.  }
\label{fig:ew}
\end{figure*}

\subsection{Application to TW~Hya}

Our imaging model assumes that the observed profile variations are mainly due to rotational modulation;  
all other sources of profile variability (and in particular intrinsic variability) cannot be properly 
reproduced and thus contribute as noise into the modelling process, degrading the imaging performance 
and potentially even drowning all relevant information.  Filtering out significant intrinsic 
variability from the observed profiles is thus worthwhile to optimise the behaviour and convergence 
of the imaging code.

We implement this by applying specific corrections on our data set.  We first suppress veiling
by scaling all LSD Stokes $I$ and $V$ photospheric profiles, to ensure that unpolarized lines have
the same equivalent widths.  We also remove the non rotationally-modulated part in the observed
fluctuations of \caii\ IRT emission, by fitting them with a sine+cosine wave (see Fig.~\ref{fig:ew})
and by scaling the
corresponding Stokes $I$ and $V$ profiles accordingly, thus ensuring that equivalent widths
of unpolarized lines match the optimal fit.  Although only approximate (especially the removal of
the intrinsic variability), this procedure is at least straightforward and has proved successful
when applied to real data \citep[e.g.,][]{Donati11} and efficient at retaining rotational modulation
mostly.  In the particular case of the present data set, the most significant effect of this 
filtering process is to remove the emission burst detected in 2008 March (at cycle 1.654 and 1.668) by 
scaling down the 2 corresponding \caii\ IRT profiles (shown as green open squares on Fig.~\ref{fig:ew} 
left panel);  the remaining modulation left in the \caii\ profiles are the $\simeq$30\% and $<10$\% 
fluctuations reported in Sec.~\ref{sec:var}.  

We use Unno-Rachkovsky's equations known to provide a good description of the local Stokes $I$ and 
$V$ profiles 
(including magneto-optical effects) in the presence of both weak and strong magnetic fields 
\citep[e.g.,][Sec.~9.8]{Landi04} despite their being based on the assumption of a simple 
Milne-Eddington atmospheric model.
The model parameters used in the specific case of TW~Hya are mostly identical to those used in our 
previous studies \citep{Donati10b, Donati11}.  
The emission profile scaling factor $\epsilon$, describing the emission
enhancement of accretion regions over the quiet chromosphere, is once again set to $\epsilon=10$.
We also assume (as in previous studies) that the magnetic topology of TW~Hya is antisymmetric 
with respect to the centre of the star;  such field configurations are indeed best suited to 
explain how accretion regions can form at high latitudes mostly \citep[e.g.,][]{Long08}, as often 
the case for cTTSs.  This assumption has little impact 
on the reconstructed magnetic maps over the visible regions of the stellar surface and allows a 
consistent and realistic (though not necessarily exact) field distribution to be recovered on 
the unseen hemisphere as well.  

\begin{figure*}
\vspace{-2mm}
\hbox{\includegraphics[scale=0.7]{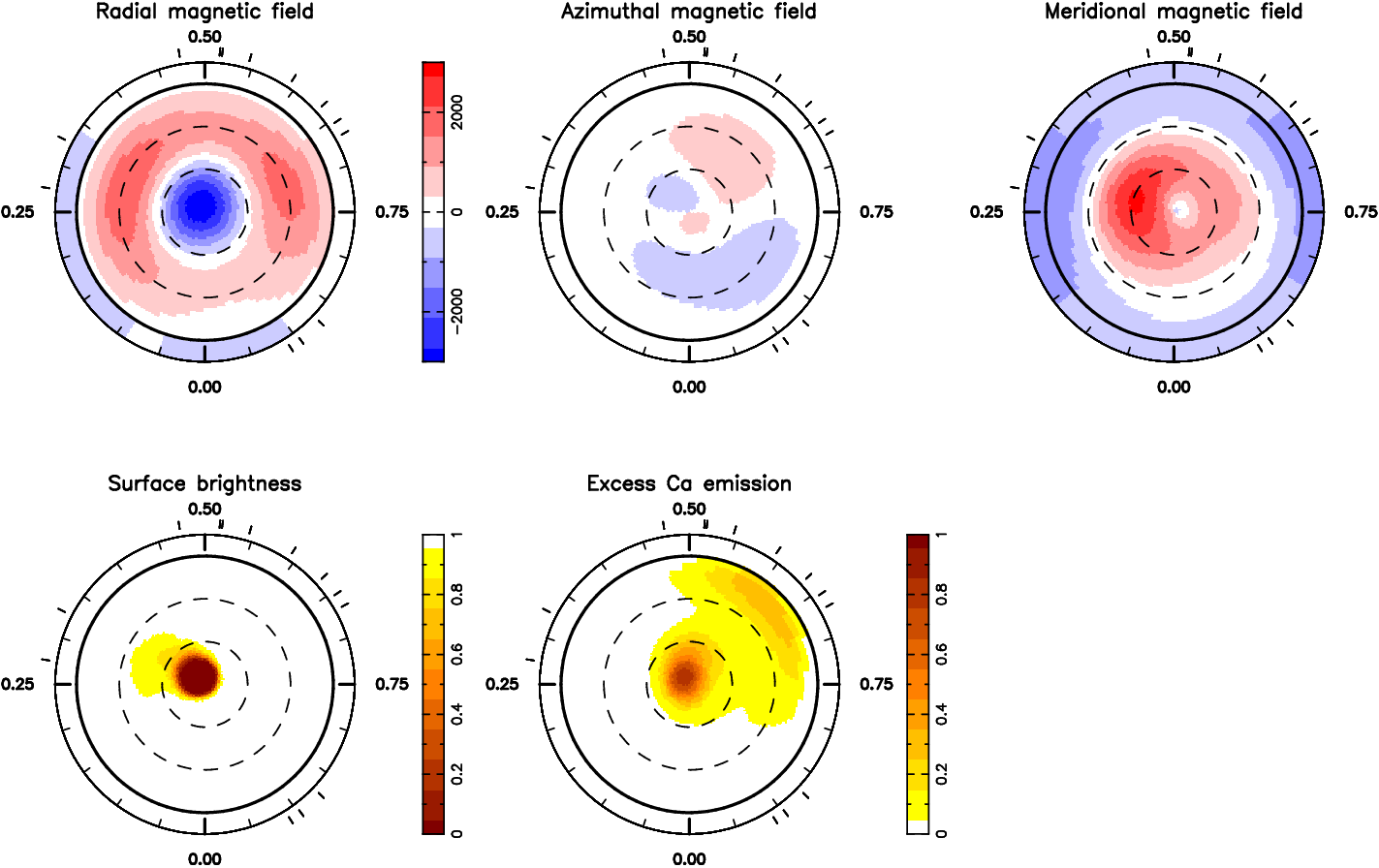}}
\vspace{8mm}
\hbox{\includegraphics[scale=0.7]{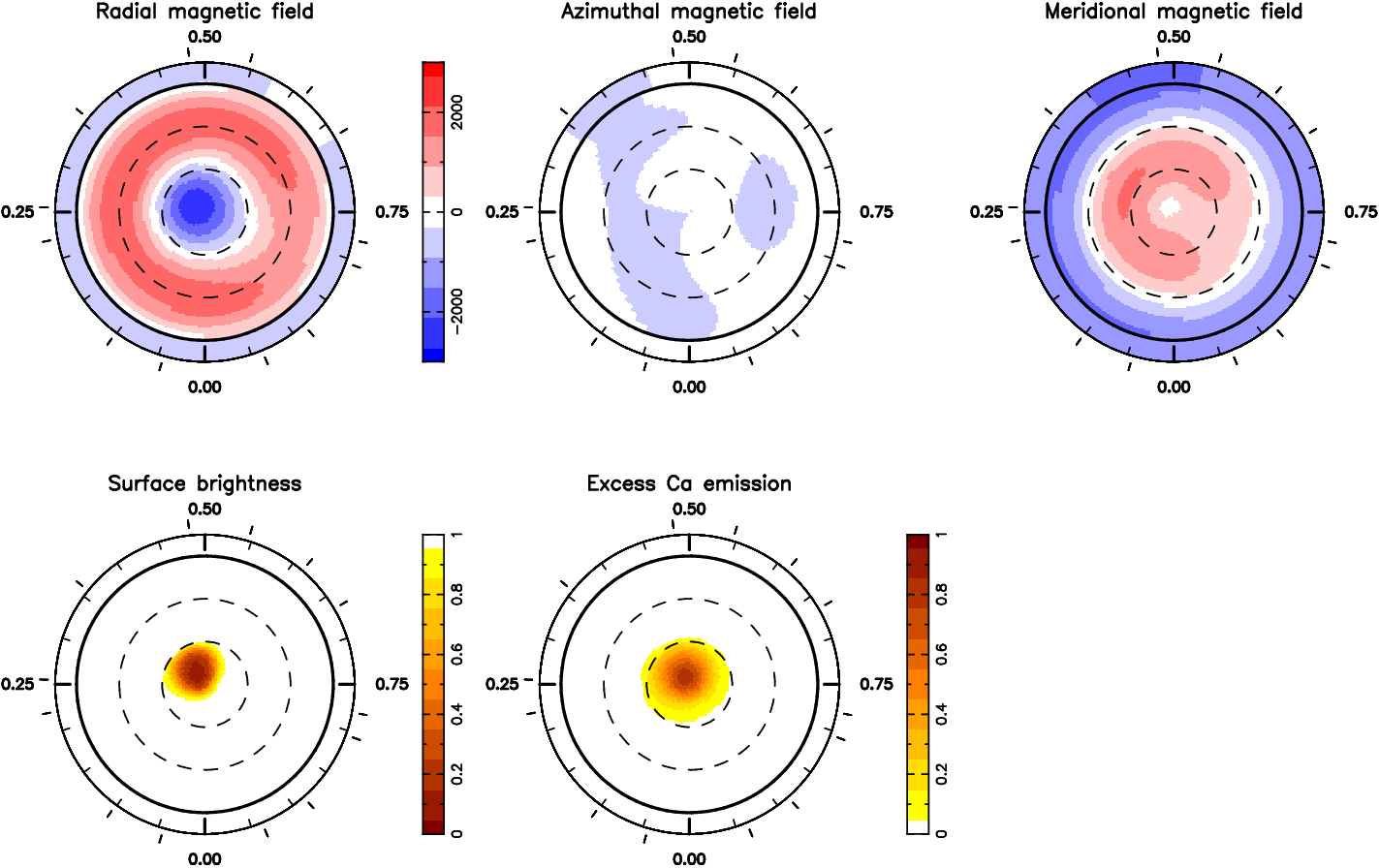}}
\caption[]{Maps of the radial, azimuthal and meridional components of the magnetic field $\bf B$
(first and third rows, left to right panels respectively), photospheric brightness and excess
\caii\ IRT emission (second and fourth rows, first and second panels respectively) at the
surface of TW~Hya, in 2008 March (top two rows) and 2010 March (bottom two rows).  
Magnetic fluxes are labelled in G;  local photospheric brightness (normalized to that of the quiet
photosphere) varies from 1 (no spot) to 0 (no light);  local excess \caii\ emission varies from 0
(no excess emission) to 1 (excess emission covering 100\% of the local grid cell, assuming an
intrinsic excess emission of 10$\times$ the quiet chromospheric emission).
In all panels, the star is shown in flattened polar projection down to latitudes of $-30\degr$,
with the equator depicted as a bold circle and parallels as dashed circles.  Radial ticks around
each plot indicate phases of observations. }  
\label{fig:map}
\end{figure*}

\begin{figure*}
\vspace{-3mm}
\center{
\includegraphics[scale=0.65,angle=-90]{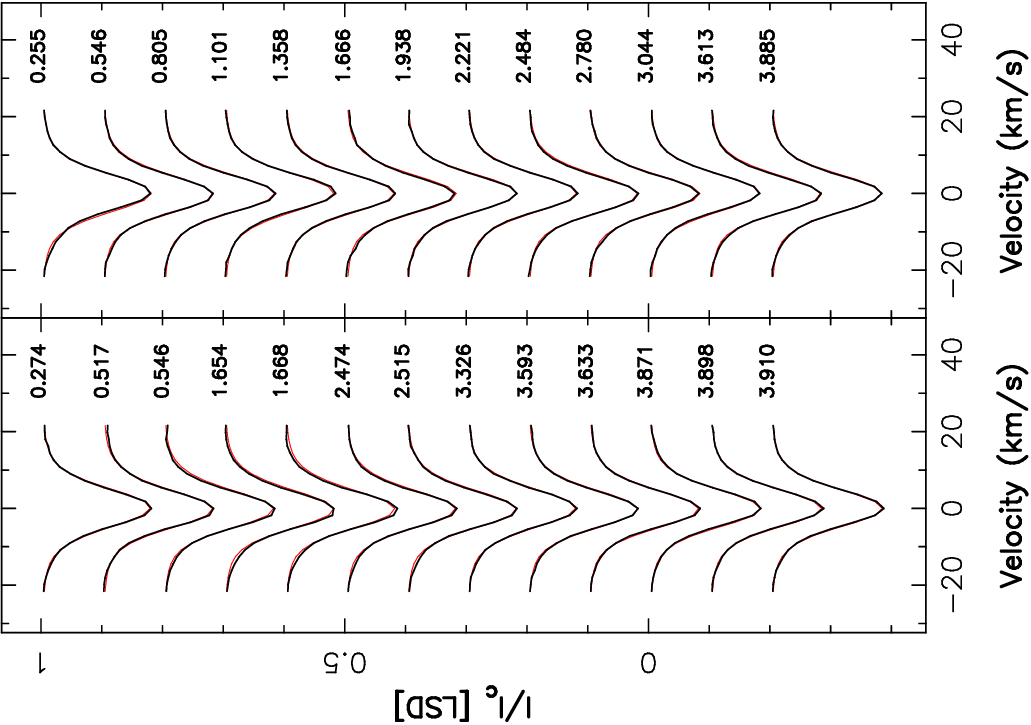}\hspace{4mm}
\includegraphics[scale=0.65,angle=-90]{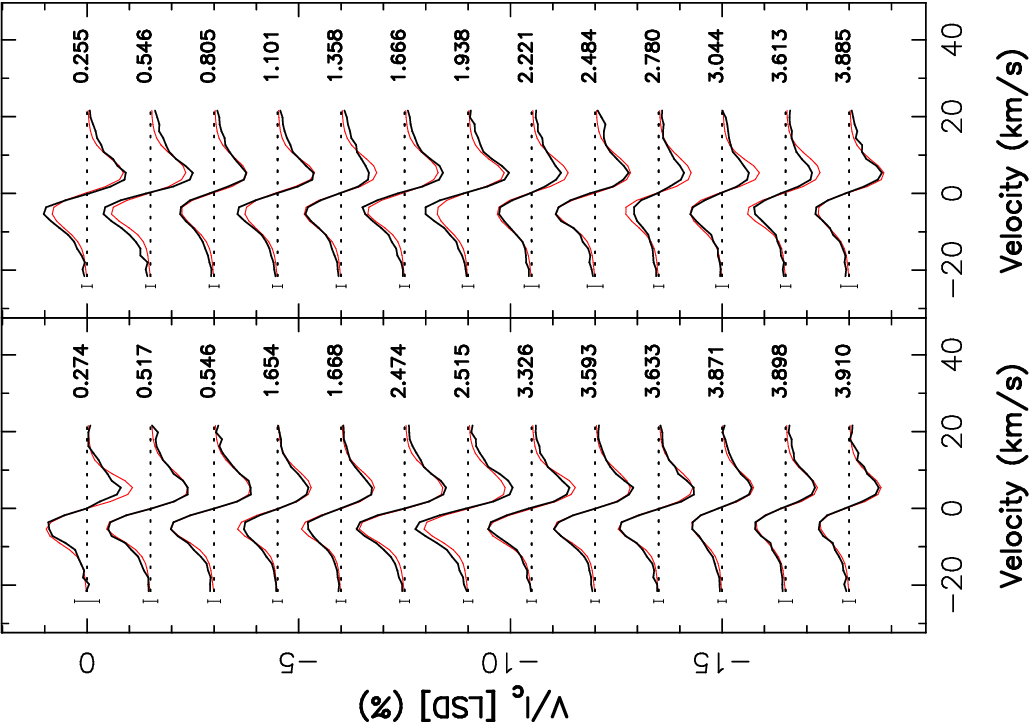}}
\vspace{3mm}
\center{
\includegraphics[scale=0.65,angle=-90]{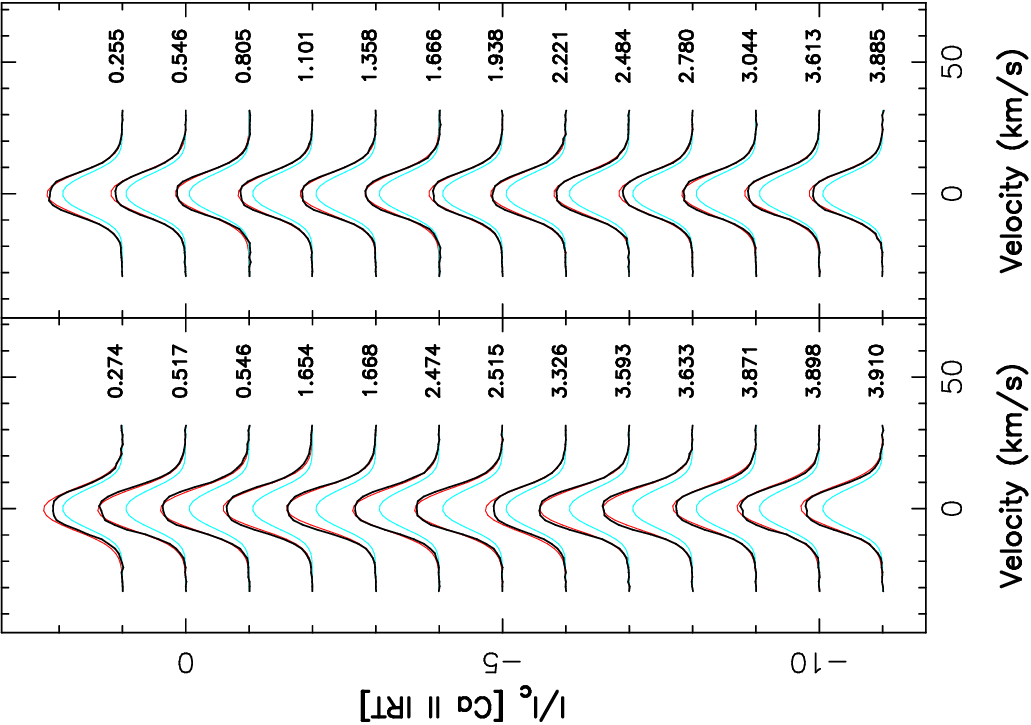}\hspace{4mm}
\includegraphics[scale=0.65,angle=-90]{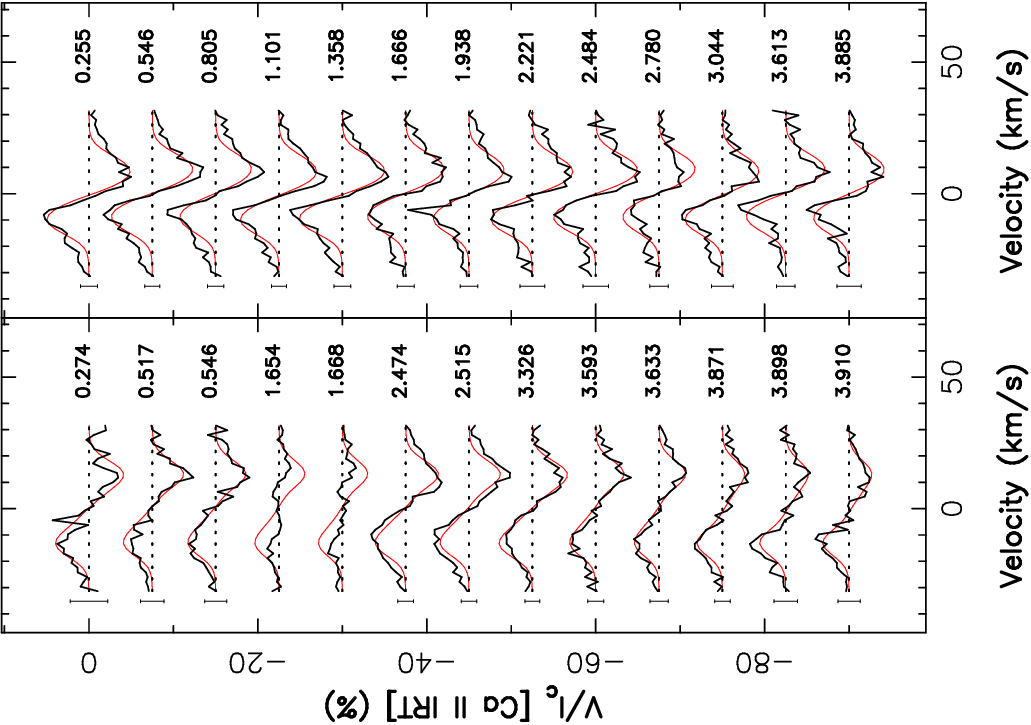}}
\caption[]{Maximum-entropy fit (thin red line) to the observed (thick black line) Stokes $I$ and
Stokes $V$ LSD photospheric profiles (top panels) and \caii\ IRT profiles (bottom panels)
of TW~Hya.  In each panel, the left and right columns correspond to the 2008 March and 2010 March 
data respectively.  The light-blue curve in the bottom left panel shows the (constant)
contribution of the quiet chromosphere to the Stokes $I$ \caii\ profiles.
Rotational cycles and 3$\sigma$ error bars (for Stokes $V$ profiles) are also shown next to each
profile (the two profiles with no error bars were excluded from the fit).  }
\label{fig:fit}
\end{figure*}

The magnetic, brightness and accretion maps we reconstruct for TW~Hya at both epochs are 
shown in Fig.~\ref{fig:map}, with the corresponding fits to the data shown in Fig.~\ref{fig:fit}.  
The SH expansion describing the field was limited to terms with $\ell\leq5$; attempts 
with $\ell=10$ indicate that little power is reconstructed in higher order modes ($\ell\ge6$), 
reflecting essentially the limited spatial information accessible to Doppler tomography at 
low \vsini\ values.  Error bars on Zeeman signatures were artificially expanded by a factor of 
3 (for LSD profiles) and 2 (for \caii\ emission) at both epochs to take into account the 
significant level of intrinsic variability;  moreover, Zeeman signatures corresponding to the 
2 \caii\ profiles recorded during the emission burst of 2008 March (at cycles 1.654 and 1.668) 
were excluded from the fit, being strongly discrepant with others taken at very similar phases 
(but different cycles, e.g., 3.633).  
The fits we finally obtain correspond to a reduced chi-square 
\chisqr\ of 1.2, starting from initial values of 59 and 47 for the 2008 March and 2010 March data 
sets respectively (with scaled-up error bars on Zeeman signatures).  

As a by-product, we obtain new estimates for various spectral characteristics of TW~Hya;  
in particular, we find that the RV is equal to $12.55\pm0.10$~\kms, that the mean red-shift 
of \caii\ emission profiles with respect to LSD profiles is $0.85\pm0.1$~\kms, and that the 
\vsini\ is $4\pm1$~\kms.  Finally, we find that the profiles are best fitted for a value of 
the local filling factor $\psi$, describing the relative proportion of magnetic areas at any
given point of the stellar surface, equal to $\simeq$0.4.  

\subsection{Modelling results}

The reconstructed large-scale magnetic topology of TW~Hya is very similar at both epochs, 
featuring a strong and mostly axisymmetric field as expected from the small amount of 
rotational modulation that Zeeman signatures exhibit.  
The magnetic field at the surface of the star has an average intensity of $\simeq$1.5~kG, 
concentrating mostly in the radial and meridional components;  in particular, the 
radial field map consists of a spot of negative radial field near the pole (where the 
field reaches up to 3~kG, in agreement with the field strengths estimated from the narrow 
\hei\ emission component) surrounded with a ring of positive radial field at low latitudes 
(see Fig.~\ref{fig:map}).  We find that the field is almost purely poloidal and mostly 
axisymmetric at both epochs, with $\leq5$\% of the reconstructed magnetic energy in the 
toroidal and non-axisymmetric (i.e., SH modes with $m>\ell/2$) poloidal fields. 

The octupolar component of the poloidal field largely dominates at both epochs (reaching 
polar strengths of 2.5 and 2.8~kG in 2008 March and 2010 March respectively) as expected 
from the topologies of the reconstructed radial and meridional field maps;  the dipole 
component is typically 4--6 times weaker, with a polar strength of 0.4 and 0.7~kG in 2008 
March and 2010 March respectively.  The octupole is aligned with respect to the rotation 
axis to better than 10\degr, with the negative pole being visible from the Earth.  
The reconstructed dipole component is also aligned with the rotation axis in 2010 March but 
tilted by about 45\degr\ (towards phase 0.5) in 2008 March.  However, given the relative 
weakness of the dipole field and the incomplete phase coverage of the dataset at this epoch, 
it is unclear whether this tilt is significant;  extensive simulations indeed suggest that 
retrieving a weak dipole component in a predominantly octupolar large-scale field is uncertain, 
especially at low \vsini's.  At both epochs, the visible pole of the dipolar component is the 
positive pole, suggesting that the dipole and octupole components are essentially anti-parallel 
in 2010 March.  Given the high-level of intrinsic variability in the data, we caution that the 
(limited) change in the large-scale magnetic topology of TW~Hya (especially regarding the dipole 
component) between the two epochs may potentially be spurious.  

The brightness distributions at the surface of TW~Hya that we recover show a cool spot near 
the pole at both epochs (see Fig.~\ref{fig:map}).  This spot, covering about 2\% of the 
stellar surface, is shifted from the pole by about 10\degr\ towards phase 0.35--0.40, as 
expected from the RV variations of photospheric LSD profiles (see Sec.~\ref{sec:var}).  
This spot is located right on the region of strong negative radial field near the pole 
(the magnetic pole of the octupole component), explaining 
a posteriori why photospheric Zeeman signatures are essentially insensitive to this strong 
negative radial field region and probe instead fields of opposite polarities (positive radial 
and meridional fields at intermediate latitudes, see Fig.~\ref{fig:map});  being very cool, 
this spot emits much less photons (per surface area) than the surrounding photosphere and 
therefore much less polarised photons as well, making it essentially invisible to Zeeman 
signatures of photospheric LSD profiles (at least at visible wavelengths).  

The distributions of accretion-powered excess emission also show at both epochs a clear spot 
near the pole, and covering 2--3\% of the stellar surface (see Fig.~\ref{fig:map}).  This 
accretion region approximately overlaps with the cool photospheric spot, with possibly a small 
phase shift (towards phase 0.50 in 2010 March in particular) suggesting that the accretion 
region may slightly trail the photospheric spot.  These distributions explain a posteriori 
why Zeeman signatures of \caii\ (and \hei) emission profiles mostly probe fields with opposite 
polarities than those traced by LSD photospheric profiles;  while \caii\ (and to a much larger 
extent \hei) emission profiles are mostly sensitive to the polar regions (hosting strong negative radial 
fields), LSD photospheric profiles provide a complementary view of the visible stellar surface 
(and in particular of intermediate latitude regions, hosting positive radial and meridional 
fields).  

The near-polar accretion region apparently also extends to low latitudes at phase $\simeq$0.6 
in 2008 March, potentially indicating that sporadic non-polar accretion is also occurring at 
times on TW~Hya.  This low-latitude appendage directly reflects the $\simeq$30\% full-amplitude 
modulation observed in the emission flux of \caii\ (and of \hei) profiles peaking around phase 0.6 
(see Sec.~\ref{sec:var}) and is consistent with veiling estimates (also peaking around 
phase 0.6 at this epoch).  Independent confirmation that this non-polar accretion event is real -- 
rather than caused by over-interpreting intrinsic variability into rotational modulation -- comes 
from the weird Zeeman signatures of \caii\ (and \hei) lines collected during the emission burst 
of 2008 March (at cycles 1.654 and 1.668) and excluded from the 
fitted data set (given their very specific shapes, see Fig.~\ref{fig:fit}).  By subtracting 
the expected Zeeman signatures (the red lines on Fig.~\ref{fig:fit}) from the observed ones 
(the black lines) for these two profiles, we end up with the Zeeman signatures associated with 
the excess emission specific to the emission burst and find that they trace a radial field 
of opposite (i.e., positive) polarity than the main Zeeman signatures collected at similar 
phases but outside of the accretion burst (e.g., at cycle 3.633).  This unambiguously demonstrates 
that sporadic accretion is truly occurring on TW~Hya in regions of positive radial field, i.e., at 
much lower latitudes than where most of the disc material is accreted (on the pole), therefore 
validating the reality of the low-latitude appendage to the near-polar accretion region in 2008 March.

\section{Summary \& discussion}
\label{sec:dis}

We present in this paper new results on magnetospheric accretion processes taking place at the 
surfaces of forming Sun-like stars;  towards this aim, we use spectropolarimetric data collected on 
the evolved cTTS TW~Hya at 2 different epochs (2008 March and 2010 March) and with 
ESPaDOnS@CFHT in the framework of the MaPP program.  
Strong Zeeman signatures are detected in the LSD profiles of photospheric lines of TW~Hya, as well 
as from the \caii\ and \hei\ emission lines commonly used as accretion proxies, revealing the 
presence of longitudinal fields ranging from --3.5 to 0.7~kG depending on the set of lines and on 
the observing epoch;  in particular, longitudinal fields of opposite polarities are detected (at 
both epochs) from the two sets of lines (i.e., LSD profiles and accretion proxies), clearly 
indicating that they probe different regions of the stellar surface.  
Using our dedicated tomographic imaging tool \citep[tested and validated with several similar 
studies, e.g.,][among the most recent ones]{Donati10b, Donati11}, we convert these data sets into images 
of the large-scale magnetic topology, as well as distributions of the photospheric brightness 
and of the accretion-powered excess emission, at the surface of TW~Hya. 

We find that the large-scale magnetic field of TW~Hya is strong (with a typical average surface 
intensity of 1.5~kG) and mostly poloidal and axisymmetric (with respect to the rotation axis).  
More specifically, the poloidal field is dominated by a strong octupole component (with a polar 
strength of 2.5--2.8~kG) mostly aligned with the rotation axis (to better than 10\degr) and 
exhibiting its negative pole to an Earth-based observer.  The dipole component is much weaker, 
by typically a factor of 4--6 (with a polar strength ranging from $\simeq$400~G in 2008 March 
to $\simeq$700~G in 2010 March), and is found to be mostly anti-parallel with respect to the 
octupole (though apparently tilted at about 45\degr\ with respect to the rotation axis in 2008 March).  
We however caution that the latter results need further confirmation from additional data 
given the fairly strong octupole to dipole intensity contrast, and the significant level of 
intrinsic variability in the data.  

Regarding its large-scale magnetic topology, TW~Hya 
is fairly similar to V2129~Oph, also hosting a strong and mostly poloidal, axisymmetric 
large-scale field in which the octupole component largely dominates \citep[][though with 
parallel, rather than antiparallel, dipole and octupole moments]{Donati11};  it 
is however significantly different from AA~Tau and BP~Tau on the one hand \citep[also hosting 
a strong, mostly axisymmetric, large-scale magnetic field, but with a larger and sometimes 
dominant dipole component,][]{Donati10b} and V4046~Sgr, CV~Cha or CR~Cha on the other hand 
\citep[exhibiting a much weaker, mostly non-axisymmetric poloidal field,][]{Hussain09, Donati11b}.  
This is in good agreement with the overall trends on the properties of dynamo fields as 
derived from previous studies, both on cTTSs and main sequence dwarfs \citep[e.g.,][]{Morin08b, 
Donati09, Morin10};  more specifically, these trends show that cool stars (i) are capable of 
producing intense axisymmetric large-scale fields with strong dipole components when fully convective 
(as for AA~Tau and BP~Tau), (ii) start losing their large-scale dipole while keeping a strong 
axisymmetric octupole when they start developing a radiative core (as for V2129~Oph and TW~Hya) 
and (iii) finally end up with weak, mostly non-axisymmetric poloidal fields when their radiative 
cores occupy a large-enough proportion of the star (with an outer radius of at least 
0.4--0.5~\rstar, as for V4046~Sgr, CV~Cha and CR~Cha).  

We also confirm that TW~Hya hosts a cool photospheric spot near the pole, as initially 
proposed by \citet{Huelamo08} to account for the observed regular RV modulation of its 
spectrum;  we find that this spot is located near the pole, shifted (by about 10\degr) 
towards phase 0.35--0.40 (in the ephemeris of Eq.~\ref{eq:eph}) and apparently very 
stable on timescales of several years.  This spot mostly overlaps with the main magnetic 
pole, which thus becomes essentially invisible to LSD photospheric profiles.  Most likely, 
this near-polar, long-lived cool spot is a direct consequence of the strong large-scale 
magnetic topology, with the intense fields (of $\simeq$3~kG) present at the magnetic pole 
presumably inhibiting outward energy transport by convection more than everywhere else at 
the surface of the star.  In this respect as well, TW~Hya resembles V2129~Oph which 
also hosts a similar cool photospheric spot near the pole \citep{Donati11}.  

We finally obtain that TW~Hya hosts a near-polar region of accretion-powered excess 
\caii\ (and \hei) emission, mostly overlapping with the cool photospheric spot and coinciding 
with the main magnetic pole.  This confirms in particular that accretion occurs mostly 
poleward at the surface of TW~Hya, again similar to what is reported for V2129~Oph 
\citep{Donati11}.  In addition to this regular poleward accretion, we also report the 
detection of sporadic accretion events towards lower latitudes.  These episodic events can 
show up as enhanced rotational modulation in \caii\ and \hei\ emission (as in our 2008 
March images);  they can even manifest themselves at times as short-lived flaring-like 
emission bursts (as in cycles 1.654 and 1.668) producing specific Zeeman signatures of 
opposite polarities than (and thus partially cancelling out with) those associated with the 
main polar accretion region (octupoles exhibiting positive and negative radial fields at low 
and high latitudes respectively, or vice versa, within a given hemisphere).  We note that 
the epoch at which equator-ward accretion occurs (2008 March) is also that at which the 
large-scale dipole field is weakest and potentially inclined with respect to the rotation 
axis (towards the phase of maximum \caii\ emission approximately);  although this needs 
confirmation from new data with better phase sampling than our 2008 March data set, it 
is at least consistent with theoretical expectations \citep[e.g.,][]{Romanova11}.  

From the emission fluxes observed in spectral lines commonly used as accretion proxies, 
we conclude that the logarithmic mass accretion rate at the surface of TW~Hya is equal to 
$-8.9\pm0.4$ (in \mspy) at both epochs, in reasonable agreement with recent independent 
determinations from optical spectra \citep[e.g.,][]{Curran11}.  
Following the theoretical work of \citet{Bessolaz08}, we can infer that TW~Hya is capable 
of magnetically disrupting its disc up to a distance \rmag\ of about 
$4\pm1$~\rstar\ if the large-scale dipole is as strong as 700~G, but only $3\pm1$~\rstar\ 
if the dipole strength is only 400~G.  In the latter case (corresponding to our 2008 March 
observations), this is just barely beyond the distance at which the dipole field starts 
to dominate over the octupole field \citep[ranging from 1.7 to 2.2~\rstar\ for octupole to 
dipole polar strength ratios of 4--6, e.g.,][]{Gregory10};  this potentially explains why 
equator-ward accretion events are observed preferentially at this epoch.  When compared 
to the corotation radius $\rcor\simeq8.3$~\rstar\ (at which the Keplerian period equals 
the stellar rotation period), we find that $\rmag/\rcor$ potentially ranges from 
$0.35\pm0.10$ to $0.50\pm0.10$ depending on the dipole field strength, far below the 
value (of $\simeq1$) at which star/disc magnetic coupling can start inducing a significant outflow 
through a propeller-like mechanism \citep[e.g.,][]{Romanova04} and thus force the star 
to spin down and rotate as slowly as one cycle per week.  

We thus speculate that TW~Hya, with its relatively short rotation period (of $\simeq$3.56~d), 
is in a phase of rapid spin-up, mostly reflecting the fact that its large-scale dipole field 
is no longer strong enough to disrupt the disc beyond 0.5~\rcor\ and fuel the associated leak 
of angular momentum (e.g., through a propeller-like mechanism);  this situation is similar, 
though admittedly not exactly identical, to that described in the recent MHD simulations of 
\citet{Zanni09}.  In this respect, TW~Hya is different from V2129~Oph whose rotation period 
(of $\simeq$6.53~d) is still rather long.  We propose that TW~Hya and V2129~Oph are both 
undergoing a process of rapid spin-up;  however, while V2129~Oph is only at the beginning of this phase,  
TW~Hya is in a much more advanced state (suggesting at the same time that TW~Hya has been 
partly convective for a longer timespan than V2129~Oph).  

As MaPP observations accumulate and are being analysed, we are progressively building a consistent 
description of what dynamo magnetic fields of cTTSs look like and how they impact the formation 
process, in particular regarding the way magnetospheric accretion processes can succeed in 
slowing down the rotation of cTTSs.  More observations are obviously needed to validate more firmly 
and extend further the initial conclusions drawn from the existing MaPP studies including this 
one.  MaPP is still ongoing, with time granted on ESPaDOnS@CFHT (and NARVAL@TBL) up to semester 
2012b;  potential extensions, aimed at carrying out a longer survey for a few specific cTTSs and/or 
at expanding the sample to wTTSs, are under study at the moment.

\section*{Acknowledgements}
We thank the referee for a detailed reading of the manuscript.  This paper is based on
observations obtained at the Canada-France-Hawaii Telescope (CFHT), operated by the National
Research Council of Canada, the Institut National des Sciences de l'Univers of the Centre
National de la Recherche Scientifique of France and the University of Hawaii.
The ``Magnetic Protostars and Planets'' (MaPP) project is supported by the
funding agencies of CFHT and TBL (through the allocation of telescope time)
and by CNRS/INSU in particular, as well as by the French ``Agence Nationale
pour la Recherche'' (ANR).

\bibliography{twhya}

\begin{thebibliography}{}

\bibitem[\protect\citeauthoryear{{Akeson}, {Millan-Gabet}, {Ciardi}, {Boden},
  {Sargent}, {Monnier}, {McAlister}, {ten Brummelaar}, {Sturmann}, {Sturmann}
  \& {Turner}}{{Akeson} et~al.}{2011}]{Akeson11}
{Akeson} R.~L.,  {Millan-Gabet} R.,  {Ciardi} D.~R.,  {Boden} A.~F.,  {Sargent}
  A.~I.,  {Monnier} J.~D.,  {McAlister} H.,  {ten Brummelaar} T.,  {Sturmann}
  J.,  {Sturmann} L.,    {Turner} N.,  2011, \apj, 728, 96

\bibitem[\protect\citeauthoryear{{Alencar} \& {Batalha}}{{Alencar} \&
  {Batalha}}{2002}]{Alencar02}
{Alencar} S.~H.~P.,  {Batalha} C.,  2002, \apj, 571, 378

\bibitem[\protect\citeauthoryear{{Andr{\'e}}, {Basu} \& {Inutsuka}}{{Andr{\'e}}
  et~al.}{2009}]{Andre09}
{Andr{\'e}} P.,  {Basu} S.,    {Inutsuka} S.,  2009, {The formation and
  evolution of prestellar cores}.
Cambridge University Press, pp 254

\bibitem[\protect\citeauthoryear{{Bessell}, {Castelli} \& {Plez}}{{Bessell}
  et~al.}{1998}]{Bessell98}
{Bessell} M.~S.,  {Castelli} F.,    {Plez} B.,  1998, \aap, 333, 231

\bibitem[\protect\citeauthoryear{{Bessolaz}, {Zanni}, {Ferreira}, {Keppens} \&
  {Bouvier}}{{Bessolaz} et~al.}{2008}]{Bessolaz08}
{Bessolaz} N.,  {Zanni} C.,  {Ferreira} J.,  {Keppens} R.,    {Bouvier} J.,
  2008, \aap, 478, 155

\bibitem[\protect\citeauthoryear{{Bouvier}, {Alencar}, {Boutelier}, {Dougados},
  {Balog}, {Grankin}, {Hodgkin}, {Ibrahimov}, {Kun}, {Magakian} \&
  {Pinte}}{{Bouvier} et~al.}{2007b}]{Bouvier07b}
{Bouvier} J.,  {Alencar} S.~H.~P.,  {Boutelier} T.,  {Dougados} C.,  {Balog}
  Z.,  {Grankin} K.,  {Hodgkin} S.~T.,  {Ibrahimov} M.~A.,  {Kun} M.,
  {Magakian} T.~Y.,    {Pinte} C.,  2007b, \aap, 463, 1017

\bibitem[\protect\citeauthoryear{{Bouvier}, {Alencar}, {Harries}, {Johns-Krull}
  \& {Romanova}}{{Bouvier} et~al.}{2007a}]{Bouvier07}
{Bouvier} J.,  {Alencar} S.~H.~P.,  {Harries} T.~J.,  {Johns-Krull} C.~M.,
  {Romanova} M.~M.,  2007a, in {Reipurth} B.,  {Jewitt} D.,   {Keil} K.,  eds,
  Protostars and Planets V {Magnetospheric Accretion in Classical T Tauri
  Stars}.
pp 479--494

\bibitem[\protect\citeauthoryear{{Brickhouse}, {Cranmer}, {Dupree}, {Luna} \&
  {Wolk}}{{Brickhouse} et~al.}{2010}]{Brickhouse10}
{Brickhouse} N.~S.,  {Cranmer} S.~R.,  {Dupree} A.~K.,  {Luna} G.~J.~M.,
  {Wolk} S.,  2010, \apj, 710, 1835

\bibitem[\protect\citeauthoryear{{Calvet}, {D'Alessio}, {Hartmann}, {Wilner},
  {Walsh} \& {Sitko}}{{Calvet} et~al.}{2002}]{Calvet02}
{Calvet} N.,  {D'Alessio} P.,  {Hartmann} L.,  {Wilner} D.,  {Walsh} A.,
  {Sitko} M.,  2002, \apj, 568, 1008

\bibitem[\protect\citeauthoryear{{Cieza}, {Schreiber}, {Romero}, {Mora},
  {Merin}, {Swift}, {Orellana}, {Williams}, {Harvey} \& {Evans}}{{Cieza}
  et~al.}{2010}]{Cieza10}
{Cieza} L.~A.,  {Schreiber} M.~R.,  {Romero} G.~A.,  {Mora} M.~D.,  {Merin} B.,
   {Swift} J.~J.,  {Orellana} M.,  {Williams} J.~P.,  {Harvey} P.~M.,
  {Evans} N.~J.,  2010, \apj, 712, 925

\bibitem[\protect\citeauthoryear{{Curran}, {Argiroffi}, {Sacco}, {Orlando},
  {Peres}, {Reale} \& {Maggio}}{{Curran} et~al.}{2011}]{Curran11}
{Curran} R.~L.,  {Argiroffi} C.,  {Sacco} G.~G.,  {Orlando} S.,  {Peres} G.,
  {Reale} F.,    {Maggio} A.,  2011, \aap, 526, A104+

\bibitem[\protect\citeauthoryear{{Donati}, {Bouvier}, {Walter}, {Gregory},
  {Skelly}, {Hussain}, {Flaccomio}, {Argiroffi}, {Grankin}, {Jardine},
  {M{\'e}nard}, {Dougados} \& {Romanova}}{{Donati} et~al.}{2011a}]{Donati11}
{Donati} J.,  {Bouvier} J.,  {Walter} F.~M.,  {Gregory} S.~G.,  {Skelly} M.~B.,
   {Hussain} G.~A.~J.,  {Flaccomio} E.,  {Argiroffi} C.,  {Grankin} K.~N.,
  {Jardine} M.~M.,  {M{\'e}nard} F.,  {Dougados} C.,    {Romanova} M.~M.,
  2011a, \mnras, 412, 2454

\bibitem[\protect\citeauthoryear{{Donati}, {Gregory}, {Montmerle}, {Maggio},
  {Argiroffi}, {Sacco}, {Hussain}, {Kastner}, {Alencar}, {Audard}, {Bouvier},
  {Damiani}, {G\"udel}, {Huenemoerder} \& {Wade}}{{Donati}
  et~al.}{2011b}]{Donati11b}
{Donati} J.,  {Gregory} S.~G.,  {Montmerle} T.,  {Maggio} A.,  {Argiroffi} C.,
  {Sacco} G.,  {Hussain} G.~A.~J.,  {Kastner} J.,  {Alencar} S.,  {Audard} M.,
  {Bouvier} J.,  {Damiani} F.,  {G\"udel} M.,  {Huenemoerder} D.,    {Wade} G.,
   2011b, \mnras, submitted

\bibitem[\protect\citeauthoryear{{Donati} \& {Landstreet}}{{Donati} \&
  {Landstreet}}{2009}]{Donati09}
{Donati} J.,  {Landstreet} J.~D.,  2009, \araa, 47, 333

\bibitem[\protect\citeauthoryear{{Donati}, {Skelly}, {Bouvier}, {Gregory},
  {Grankin}, {Jardine}, {Hussain}, {M{\'e}nard}, {Dougados}, {Unruh},
  {Mohanty}, {Auri{\`e}re}, {Morin} \& {Far{\`e}s}}{{Donati}
  et~al.}{2010}]{Donati10b}
{Donati} J.,  {Skelly} M.~B.,  {Bouvier} J.,  {Gregory} S.~G.,  {Grankin}
  K.~N.,  {Jardine} M.~M.,  {Hussain} G.~A.~J.,  {M{\'e}nard} F.,  {Dougados}
  C.,  {Unruh} Y.,  {Mohanty} S.,  {Auri{\`e}re} M.,  {Morin} J.,
  {Far{\`e}s} R.,  2010, \mnras, 409, 1347

\bibitem[\protect\citeauthoryear{{Donati}}{{Donati}}{2003}]{Donati03}
{Donati} J.-F.,  2003, in {Trujillo-Bueno} J.,  {Sanchez Almeida} J.,  eds,
  Astronomical Society of the Pacific Conference Series Vol.~307 of
  Astronomical Society of the Pacific Conference Series, {ESPaDOnS: An Echelle
  SpectroPolarimetric Device for the Observation of Stars at CFHT}.
pp 41

\bibitem[\protect\citeauthoryear{{Donati}, {Howarth}, {Jardine}, {Petit},
  {Catala}, {Landstreet}, {Bouret}, {Alecian}, {Barnes}, {Forveille}, {Paletou}
  \& {Manset}}{{Donati} et~al.}{2006}]{Donati06b}
{Donati} J.-F.,  {Howarth} I.~D.,  {Jardine} M.~M.,  {Petit} P.,  {Catala} C.,
  {Landstreet} J.~D.,  {Bouret} J.-C.,  {Alecian} E.,  {Barnes} J.~R.,
  {Forveille} T.,  {Paletou} F.,    {Manset} N.,  2006, \mnras, 370, 629

\bibitem[\protect\citeauthoryear{{Donati}, {Jardine}, {Gregory}, {Petit},
  {Bouvier}, {Dougados}, {M{\'e}nard}, {Cameron}, {Harries}, {Jeffers} \&
  {Paletou}}{{Donati} et~al.}{2007}]{Donati07}
{Donati} J.-F.,  {Jardine} M.~M.,  {Gregory} S.~G.,  {Petit} P.,  {Bouvier} J.,
   {Dougados} C.,  {M{\'e}nard} F.,  {Cameron} A.~C.,  {Harries} T.~J.,
  {Jeffers} S.~V.,    {Paletou} F.,  2007, \mnras, 380, 1297

\bibitem[\protect\citeauthoryear{{Donati}, {Jardine}, {Gregory}, {Petit},
  {Paletou}, {Bouvier}, {Dougados}, {M{\'e}nard}, {Cameron}, {Harries},
  {Hussain}, {Unruh}, {Morin}, {Marsden}, {Manset}, {Auri{\`e}re}, {Catala} \&
  {Alecian}}{{Donati} et~al.}{2008b}]{Donati08}
{Donati} J.-F.,  {Jardine} M.~M.,  {Gregory} S.~G.,  {Petit} P.,  {Paletou} F.,
   {Bouvier} J.,  {Dougados} C.,  {M{\'e}nard} F.,  {Cameron} A.~C.,  {Harries}
  T.~J.,  {Hussain} G.~A.~J.,  {Unruh} Y.,  {Morin} J.,  {Marsden} S.~C.,
  {Manset} N.,  {Auri{\`e}re} M.,  {Catala} C.,    {Alecian} E.,  2008b, \mnras,
  386, 1234

\bibitem[\protect\citeauthoryear{{Donati}, {Moutou}, {Far{\`e}s}, {Bohlender},
  {Catala}, {Deleuil}, {Shkolnik}, {Cameron}, {Jardine} \& {Walker}}{{Donati}
  et~al.}{2008a}]{Donati08b}
{Donati} J.-F.,  {Moutou} C.,  {Far{\`e}s} R.,  {Bohlender} D.,  {Catala} C.,
  {Deleuil} M.,  {Shkolnik} E.,  {Cameron} A.~C.,  {Jardine} M.~M.,    {Walker}
  G.~A.~H.,  2008a, \mnras, 385, 1179

\bibitem[\protect\citeauthoryear{{Donati}, {Semel}, {Carter}, {Rees} \&
  {Collier Cameron}}{{Donati} et~al.}{1997}]{Donati97b}
{Donati} J.-F.,  {Semel} M.,  {Carter} B.~D.,  {Rees} D.~E.,    {Collier
  Cameron} A.,  1997, \mnras, 291, 658

\bibitem[\protect\citeauthoryear{{Fang}, {van Boekel}, {Wang}, {Carmona},
  {Sicilia-Aguilar} \& {Henning}}{{Fang} et~al.}{2009}]{Fang09}
{Fang} M.,  {van Boekel} R.,  {Wang} W.,  {Carmona} A.,  {Sicilia-Aguilar} A.,
    {Henning} T.,  2009, \aap, 504, 461

\bibitem[\protect\citeauthoryear{{Gregory}, {Jardine}, {Gray} \&
  {Donati}}{{Gregory} et~al.}{2010}]{Gregory10}
{Gregory} S.~G.,  {Jardine} M.,  {Gray} C.~G.,    {Donati} J.,  2010, Reports
  on Progress in Physics, 73, 126901

\bibitem[\protect\citeauthoryear{{G{\"u}nther}, {Schmitt}, {Robrade} \&
  {Liefke}}{{G{\"u}nther} et~al.}{2007}]{Gunther07b}
{G{\"u}nther} H.~M.,  {Schmitt} J.~H.~M.~M.,  {Robrade} J.,    {Liefke} C.,
  2007, \aap, 466, 1111

\bibitem[\protect\citeauthoryear{{Hu{\'e}lamo}, {Figueira}, {Bonfils},
  {Santos}, {Pepe}, {Gillon}, {Azevedo}, {Barman}, {Fern{\'a}ndez}, {di Folco},
  {Guenther}, {Lovis}, {Melo}, {Queloz} \& {Udry}}{{Hu{\'e}lamo}
  et~al.}{2008}]{Huelamo08}
{Hu{\'e}lamo} N.,  {Figueira} P.,  {Bonfils} X.,  {Santos} N.~C.,  {Pepe} F.,
  {Gillon} M.,  {Azevedo} R.,  {Barman} T.,  {Fern{\'a}ndez} M.,  {di Folco}
  E.,  {Guenther} E.~W.,  {Lovis} C.,  {Melo} C.~H.~F.,  {Queloz} D.,    {Udry}
  S.,  2008, \aap, 489, L9

\bibitem[\protect\citeauthoryear{{Hussain}, {Collier Cameron}, {Jardine},
  {Dunstone}, {Velez}, {Stempels}, {Donati}, {Semel}, {Aulanier}, {Harries},
  {Bouvier}, {Dougados}, {Ferreira}, {Carter} \& {Lawson}}{{Hussain}
  et~al.}{2009}]{Hussain09}
{Hussain} G.~A.~J.,  {Collier Cameron} A.,  {Jardine} M.~M.,  {Dunstone} N.,
  {Velez} J.~R.,  {Stempels} H.~C.,  {Donati} J.-F.,  {Semel} M.,  {Aulanier}
  G.,  {Harries} T.,  {Bouvier} J.,  {Dougados} C.,  {Ferreira} J.,  {Carter}
  B.~D.,    {Lawson} W.~A.,  2009, \mnras, pp 997

\bibitem[\protect\citeauthoryear{{Johns-Krull}}{{Johns-Krull}}{2007}]{Johns07}
{Johns-Krull} C.~M.,  2007, \apj, 664, 975

\bibitem[\protect\citeauthoryear{{Kastner}, {Huenemoerder}, {Schulz},
  {Canizares} \& {Weintraub}}{{Kastner} et~al.}{2002}]{Kastner02}
{Kastner} J.~H.,  {Huenemoerder} D.~P.,  {Schulz} N.~S.,  {Canizares} C.~R.,
  {Weintraub} D.~A.,  2002, \apj, 567, 434

\bibitem[\protect\citeauthoryear{{Kurucz}}{{Kurucz}}{1993}]{Kurucz93}
{Kurucz} R.,  1993, CDROM \#~13 (ATLAS9 atmospheric models) and \#~18 (ATLAS9
  and SYNTHE routines, spectral line database).
Smithsonian Astrophysical Observatory, Washington D.C.

\bibitem[\protect\citeauthoryear{{Landi degl'Innocenti} \& {Landolfi}}{{Landi
  degl'Innocenti} \& {Landolfi}}{2004}]{Landi04}
{Landi degl'Innocenti} E.,  {Landolfi} M.,  2004, {Polarisation in spectral
  lines}.
Dordrecht/Boston/London: Kluwer Academic Publishers

\bibitem[\protect\citeauthoryear{{Long}, {Romanova} \& {Lovelace}}{{Long}
  et~al.}{2008}]{Long08}
{Long} M.,  {Romanova} M.~M.,    {Lovelace} R.~V.~E.,  2008, \mnras, 386, 1274

\bibitem[\protect\citeauthoryear{{Morin}, {Donati}, {Petit}, {Delfosse},
  {Forveille} \& {Jardine}}{{Morin} et~al.}{2010}]{Morin10}
{Morin} J.,  {Donati} J.,  {Petit} P.,  {Delfosse} X.,  {Forveille} T.,
  {Jardine} M.~M.,  2010, \mnras, 407, 2269

\bibitem[\protect\citeauthoryear{{Morin}, {Donati}, {Petit}, {Delfosse},
  {Forveille}, {Albert}, {Auri{\`e}re}, {Cabanac}, {Dintrans}, {Fares},
  {Gastine}, {Jardine}, {Ligni{\`e}res}, {Paletou}, {Ramirez Velez} \&
  {Th{\'e}ado}}{{Morin} et~al.}{2008}]{Morin08b}
{Morin} J.,  {Donati} J.-F.,  {Petit} P.,  {Delfosse} X.,  {Forveille} T.,
  {Albert} L.,  {Auri{\`e}re} M.,  {Cabanac} R.,  {Dintrans} B.,  {Fares} R.,
  {Gastine} T.,  {Jardine} M.~M.,  {Ligni{\`e}res} F.,  {Paletou} F.,  {Ramirez
  Velez} J.~C.,    {Th{\'e}ado} S.,  2008, \mnras, 390, 567

\bibitem[\protect\citeauthoryear{{Natta}, {Testi}, {Muzerolle}, {Randich},
  {Comer{\'o}n} \& {Persi}}{{Natta} et~al.}{2004}]{Natta04}
{Natta} A.,  {Testi} L.,  {Muzerolle} J.,  {Randich} S.,  {Comer{\'o}n} F.,
  {Persi} P.,  2004, \aap, 424, 603

\bibitem[\protect\citeauthoryear{{Qi}, {Ho}, {Wilner}, {Takakuwa}, {Hirano},
  {Ohashi}, {Bourke}, {Zhang}, {Blake}, {Hogerheijde}, {Saito}, {Choi} \&
  {Yang}}{{Qi} et~al.}{2004}]{Qi04}
{Qi} C.,  {Ho} P.~T.~P.,  {Wilner} D.~J.,  {Takakuwa} S.,  {Hirano} N.,
  {Ohashi} N.,  {Bourke} T.~L.,  {Zhang} Q.,  {Blake} G.~A.,  {Hogerheijde} M.,
   {Saito} M.,  {Choi} M.,    {Yang} J.,  2004, \apjl, 616, L11

\bibitem[\protect\citeauthoryear{{Romanova}, {Long}, {Lamb}, {Kulkarni} \&
  {Donati}}{{Romanova} et~al.}{2011}]{Romanova11}
{Romanova} M.~M.,  {Long} M.,  {Lamb} F.~K.,  {Kulkarni} A.~K.,    {Donati} J.,
   2011, \mnras, 411, 915

\bibitem[\protect\citeauthoryear{{Romanova}, {Ustyugova}, {Koldoba} \&
  {Lovelace}}{{Romanova} et~al.}{2004}]{Romanova04}
{Romanova} M.~M.,  {Ustyugova} G.~V.,  {Koldoba} A.~V.,    {Lovelace} R.~V.~E.,
   2004, \apjl, 616, L151

\bibitem[\protect\citeauthoryear{{Rucinski} \& {Krautter}}{{Rucinski} \&
  {Krautter}}{1983}]{Rucinski83}
{Rucinski} S.~M.,  {Krautter} J.,  1983, \aap, 121, 217

\bibitem[\protect\citeauthoryear{{Rucinski}, {Matthews}, {Kuschnig},
  {Pojma{\'n}ski}, {Rowe}, {Guenther}, {Moffat}, {Sasselov}, {Walker} \&
  {Weiss}}{{Rucinski} et~al.}{2008}]{Rucinski08}
{Rucinski} S.~M.,  {Matthews} J.~M.,  {Kuschnig} R.,  {Pojma{\'n}ski} G.,
  {Rowe} J.,  {Guenther} D.~B.,  {Moffat} A.~F.~J.,  {Sasselov} D.,  {Walker}
  G.~A.~H.,    {Weiss} W.~W.,  2008, \mnras, 391, 1913

\bibitem[\protect\citeauthoryear{{Sanchez Almeida} \& {Lites}}{{Sanchez
  Almeida} \& {Lites}}{1992}]{Sanchez92}
{Sanchez Almeida} J.,  {Lites} B.~W.,  1992, \apj, 398, 359

\bibitem[\protect\citeauthoryear{{Setiawan}, {Henning}, {Launhardt},
  {M{\"u}ller}, {Weise} \& {K{\"u}rster}}{{Setiawan} et~al.}{2008}]{Setiawan08}
{Setiawan} J.,  {Henning} T.,  {Launhardt} R.,  {M{\"u}ller} A.,  {Weise} P.,
   {K{\"u}rster} M.,  2008, \nat, 451, 38

\bibitem[\protect\citeauthoryear{{Siess}, {Dufour} \& {Forestini}}{{Siess}
  et~al.}{2000}]{Siess00}
{Siess} L.,  {Dufour} E.,    {Forestini} M.,  2000, \aap, 358, 593

\bibitem[\protect\citeauthoryear{{Thi}, {Mathews}, {M{\'e}nard}, {Woitke},
  {Meeus}, {Riviere-Marichalar}, {Pinte}, {Howard}, {Roberge}, {Sandell},
  {Pascucci}, {Riaz}, {Grady}, {Dent}, {Kamp} \& {et}}{{Thi}
  et~al.}{2010}]{Thi10}
{Thi} W.,  {Mathews} G.,  {M{\'e}nard} F.,  {Woitke} P.,  {Meeus} G.,
  {Riviere-Marichalar} P.,  {Pinte} C.,  {Howard} C.~D.,  {Roberge} A.,
  {Sandell} G.,  {Pascucci} I.,  {Riaz} B.,  {Grady} C.~A.,  {Dent} W.~R.~F.,
  {Kamp} I.,    {et} a.,  2010, \aap, 518, L125+

\bibitem[\protect\citeauthoryear{{Torres}, {Quast}, {Melo} \&
  {Sterzik}}{{Torres} et~al.}{2008}]{Torres08}
{Torres} C.~A.~O.,  {Quast} G.~R.,  {Melo} C.~H.~F.,    {Sterzik} M.~F.,  2008,
  {Young Nearby Loose Associations}.
pp 757

\bibitem[\protect\citeauthoryear{{Torres}, {Guenther}, {Marschall},
  {Neuh{\"a}user}, {Latham} \& {Stefanik}}{{Torres} et~al.}{2003}]{Torres03}
{Torres} G.,  {Guenther} E.~W.,  {Marschall} L.~A.,  {Neuh{\"a}user} R.,
  {Latham} D.~W.,    {Stefanik} R.~P.,  2003, \aj, 125, 825

\bibitem[\protect\citeauthoryear{{Yang}, {Johns-Krull} \& {Valenti}}{{Yang}
  et~al.}{2005}]{Yang05}
{Yang} H.,  {Johns-Krull} C.~M.,    {Valenti} J.~A.,  2005, \apj, 635, 466

\bibitem[\protect\citeauthoryear{{Yang}, {Johns-Krull} \& {Valenti}}{{Yang}
  et~al.}{2007}]{Yang07}
{Yang} H.,  {Johns-Krull} C.~M.,    {Valenti} J.~A.,  2007, \aj, 133, 73

\bibitem[\protect\citeauthoryear{{Zanni} \& {Ferreira}}{{Zanni} \&
  {Ferreira}}{2009}]{Zanni09}
{Zanni} C.,  {Ferreira} J.,  2009, \aap, 508, 1117

\end{thebibliography}
\bibliographystyle{mn2e}
\end{document}